\documentclass[twocolumn]{aa}
\usepackage[utf8]{inputenc}
\usepackage{amsfonts}
\usepackage{amssymb}
\usepackage{amsmath}
\usepackage{natbib}
\usepackage{graphicx}
\usepackage{txfonts}
\usepackage{textcomp}
\usepackage{subfigure}
\usepackage[pagebackref,colorlinks,citecolor=blue,urlcolor=blue,linkcolor=blue]{hyperref}
\usepackage{wasysym}
\usepackage{color}

\graphicspath{{figures/}}

\usepackage{siunitx}
\begin{document}
\title{Evidence for self-interaction of charge distribution in charge-coupled devices.}

\author{
A.~Guyonnet\inst{1} 
\and 
P.~Astier\inst{1} 
\and 
P.~Antilogus\inst{1} 
\and
N.~Regnault\inst{1} 
\and
P.~Doherty\inst{2} 
}

\institute{
LPNHE, CNRS-IN2P3 and Universit\'{e}s Paris 6 \& 7, 4 place Jussieu, F-75252 Paris Cedex 05, France
 \and 
Department of Physics, Harvard University, 17 Oxford Street, Cambridge, MA 02138, USA
}

\titlerunning{Evidence for self interaction of charge distribution in CCDs.}

\offprints{guyonnet\@@lpnhe.in2p3.fr}

\date{Received Sept 01, 2014; accepted Dec 17, 2014}

\abstract{Charge-coupled devices (CCDs) are widely used in astronomy to carry out a
  variety of measurements, such as for flux or shape of astrophysical
  objects. The data reduction procedures almost always assume that the
  response of a given pixel to illumination is independent of the
  content of the neighboring pixels. We show evidence that this simple picture
  is not exact for several CCD sensors. Namely, we provide
  evidence that localized distributions of charges (resulting from star 
  illumination or laboratory
  luminous spots) tend to broaden linearly with increasing brightness by up to
  a few percent over the whole dynamic range. We propose a physical
  explanation for this ``brighter-fatter'' effect, which implies that
  flatfields do not exactly follow Poisson statistics: the variance of
  flatfields grows less rapidly than their average, and neighboring
  pixels show covariances, which increase similarly to the square of the 
  flatfield average. These covariances decay rapidly with pixel
  separation. We observe the expected departure from Poisson statistics of
  flatfields on CCD devices and show that the observed
  effects are compatible with Coulomb forces induced by stored charges
  that deflect forthcoming charges. We extract the strength of
  the deflections from the correlations of flatfield images and derive
  the evolution of star shapes with increasing flux. We show for
  three types of sensors that within statistical uncertainties, 
  our proposed method properly bridges
  statistical properties of flatfields and the brighter-fatter
  effect.}

\maketitle

\section{Introduction} 

{Charge-coupled devices (CCDs) for astronomy are often pictured as a regular lattice of
``charge receivers'' which transform incoming light into
independent pixels in an almost linear fashion. Reduction techniques
generally assume that the response of each pixel is independent of the
charge content of its neighbors. This representation even accommodates the
non-linearity of a pixel response, typically attributed to the
electronic chain. Once linearity is restored, the data reduction
usually relies on the hypothesis that the response to a source of
light of a given shape just scales with its flux. For devices
observing the sky, the overall response of an imaging system is
described by its point spread function (PSF), which is, in practice, the
response to stars, because they are not resolved.  Astronomical
softwares (e.g., DaoPhot, PSFeX) commonly assume that star images are
homothetic and hence that bright stars can be used to account for the
impact of the whole instrument (possibly including atmosphere) on
the observed shapes of other objects, which are typically faint stars or
galaxies. For example, bright stars are also commonly used to model
the shape of faint stars, which allows to measure optimally the flux and
position of the latter with respect to shot noise.

  Do star images really scale with their flux?  Lateral diffusion of charges
during their drift in the silicon does not break this scaling, except for
an effect that we discuss later. Obviously, at some
point, the scaling
breaks down because some sort of saturation occurs in
either the sensors themselves (charge buckets overflow), or
the electronic chain (some signals saturate). However, well below saturation,
the scaling of stars images is assumed to be exact. We show 
in this paper that this is not usually true at all brightness levels and that
the effect cannot be ignored for some scientific applications.

There is a well-known hint that CCD pixels are sensitive to their
environment: the variance of flatfields does not rise linearly
with illumination, but the rise flattens out, departing from 
exact Poisson statistics (\citealt{Downing06}). 
This effect tends to vanish when one rebins the
image prior to computing the statistics, indicating that the variance
of a sum of neighboring pixels is larger than the sum of their variances.
Neighbor pixels should therefore be positively correlated (as found in
\citealt{Downing06}), and in this paper we
present direct measurements that confirm this.

If some physical phenomenon tends to reduce the variance of
flatfields, one might expect the same phenomenon to also smooth
stellar images. We show that a variety of CCD sensors deliver
stellar images that broaden with increasing flux at least to some level.
This ``brighter-fatter'' effect complicates the direct use of
stars as PSF models.

There are currently large-scale imaging programs, either underway or
planned, that intend to measure the cosmic gravitational shear induced by
mass concentrations that are located between background galaxies and observers by
evaluating the average elongation of galaxy images (e.g., \citealt{Chang12}). 
These programs
vitally rely on the shape of stars in the science images, which are used to
measure and account for the distortions induced by the observing
system on the background galaxies. It is mandatory for these programs
that the characteristics of the PSF and, in particular, its
angular size are accurately measured, typically to a fractional
accuracy of $\sim 10^{-3}$ (\citealt{BernsteinJarvis02},
\citealt{Amara21112008}). 
For the Euclid space mission, it is required that the PSF ellipticity 
is known to 0.02\%, and the PSF size to 0.1\% \citep{EuclidRB}.
Since the scaling of
stellar images is for some sensors violated at a few percent level,
bright stars cannot readily be used as models for the shape distortion
induced by the observing system (\citealt{2014arXiv1405.4285M}). 

The evolution of star shapes with
flux also adversely impacts the accuracy of PSF photometry of faint
sources. Supernova photometry for cosmology has become a demanding application
because photometric biases at the $\sim$0.006 mag level seriously
impact the precision of cosmological constraints (e.g.,
\cite{Betoule14} and references therein). Since supernova photometry
consists of measuring the PSF flux ratio of the faint
supernovae to that of bright stars around it, ignoring the
brighter-fatter effect leads to a relative flux bias equal to the PSF
size variation between these two classes of objects
(e.g., \citealt{Astier13}).
Understanding the causes of the brighter-fatter effect is
needed to accurately use the stars to model the actual PSF.

In this paper, we show that the correlations between neighbor pixels and
the brighter-fatter effect can be explained by alterations to the
drift field caused by charges already collected in the potential wells
of the CCD. However, the size of the induced distortions and how they
decay with distance from the sources both depend on manufacturing details of the CCD. 
The CCD vendors do not necessarily know these details with a precision
sufficient to model the field distortions, or even 
regard them as proprietary. In this context,
relating the statistical correlations between nearby pixels in
flatfields and the brighter-fatter effect might be the practical way
to derive the details of the latter from measurements of the
former. This paper aims to provide encouraging indications that this
program is indeed plausible. A previous communication
  (\citealt{BNL13}) presented this idea together with preliminary
  results. Here, we present a refined analysis and
  describe the matter in more detail.

We use data sets from three different instruments. 
The first is the MegaCam camera, which has been mounted on the Canada-France-Hawaii Telescope
(CFHT), since 2003. The camera is a mosaic of 36 thinned CCDs. The two other instruments are 
both equipped with deep-depletion sensors. One is the DECam camera that has been mounted on
the Blanco four-meter telescope at CTIO since 2012. DECam is a mosaic of 62 CCDs. 
The other is a CCD from E2V that is being tested by the LSST collaboration as a candidate 
for the focal plane of the future LSST telescope (\S \ref{sec:sensors}). 
For these instruments, 
we first establish the existence of both the
brighter-fatter effect (\S \ref{sec:bop}) and confirm that correlations in
flatfield exposures do exist (\S \ref{sec:ppc}). Secondly, we show how taking simple
electrostatic repulsion between collected charges and drifting electrons within 
the CCD bulk into account qualitatively describe both effects (\S
\ref{coulomb}) and that a simple model fitted to correlations in
flatfield exposures quantitatively predicts the amplitude of the brighter-fatter
effect (\S \ref{sec:comp}), thus giving a practical method to account for the variation 
of image quality (IQ) versus flux.

\section{The sensors and data}
\label{sec:sensors}

The present evolution in camera design,  moving
from using thinned CCDs to using thick CCDs, corresponds to an improvement of silicon 
wafer resistivity allowing for depletion of the mobile charge carriers
 over several hundred microns, which increases detector spatial resolution.
This work has gathered data from both types of CCDs to strengthen
the evidence that evolution of electrostatic forces within pixels 
 has detectable consequences and that it is a general feature of CCDs. 

This section describes the instruments that have been used to establish 
this hypothesis. For each instrument, the data set is constituted by both 
point source illumination exposures (using artificial spots or real stars) and by uniform 
illuminations (called hereafter flatfield illumination exposures, or simply flatfields).

\subsection{The MegaCam instrument}

Since 2003 MegaCam is a 1 deg$^2$ wide field imager hosted in the dedicated prime focus
environment, MegaPrime, on the Canada-France-Hawaii 3.6-m telescope
\citep{Boulade2003}. The camera images a field of view of ${\mathrm
  0.96 \times 0.94\ \mathrm{deg}^2}$ using 36 thinned E2V 42-90
${\mathrm 2048\times 4612}$ CCDs with \SI{13.5}{\micro\metre} pixels.  
Pixels have a conventional three-phase
structure with one electrode that defines the
collection area and the other two that constitute barriers in transfer direction.
Each CCD is read out by two amplifiers, which allows one to
read out the whole focal plane in about 35 seconds. The output of each
amplifier is sampled by a 16 bit ADC. The gains of the readout chains
have been set to $\sim 1.5\ \mathrm{e^- / ADU}$ with the consequence
that only half of the CCD full well (${\mathrm\sim 200,000\ \mathrm{e}^-}$) is
actually sampled by the readout electronics. 
The PSF broadening seen by the MegaCam instrument has been measured using all the CFHT-DEEP 
$r$-band exposures obtained during the five years of the CFHT-LS (SNLS) survey. 
The flatfields that we used are constituted by
images that were acquired using the internal illumination 
capability of the instrument (red LEDs).
This setup has been preferred from twilight flatfield because of its better reproductibity 
when acquiring pair images.

\subsection{The DECam instrument}

The DECam is a 2.2 deg$^2$ wide field imager \citep{Estrada2010}. It is mounted on
the Blanco four-meter telescope at CTIO and is used by the Dark Energy
Survey (DES) that has begun on August 31, 2013. It is made of 62 LBL/DALSA
\SI{250}{\micro\metre} thick, back-illuminated sensors \citep{holland2003}. 
Each CCD is made of 2048
$\times$ 4096 p-channel pixels, which have a width of 15$\times$15 \SI{}{\micro\metre}
and a thickness of \SI{250}{\micro\metre},
with a three-phase structure with full well at ${\mathrm\sim 200,000\ \mathrm{e}^-}$.
Charges collected in
the depletion region are stored in the buried channels that are established a
few \SI{}{\micro\metre} away from gate electrodes. Substrate
bias is 40~V so as to fully deplete the device, thus reducing charge
diffusion to a minimum (measurements indicate $\approx$ \SI{6}{\micro\metre} at 40~V). 
Each CCD is read out by two amplifiers.  

The data set used in this study comes from publicly available 
science verification images acquired during the
commissioning of the instrument between Summer 2012 
and Spring 2013 (\citealt{berntein2013}). Astronomical images are observations 
in all bands ($u$, $g$, $r$, $i$, $z$, $Y$) of three dense 
stellar fields at low galactic latitude. 
For the uniform illuminations, the analysis necessitates high stability of 
the field: we use flatfields acquired using 
 a screen installed under the dome and illuminated
by LEDs \citep{marshall2013}.

\subsection{The LSST CCD E2V-250}

The Large Synoptic Survey Telescope (LSST) will carry out a deep astronomical imaging 
survey of the southern sky. It will use an 8.4-meter ground-based telescope 
with which the construction has begun.
The survey is planned to start in 2022 and the LSST collaboration
shall soon begin the construction of the focal plane, which will 
eventually be a 9.6 deg$^2$  wide field imager made of 189 CCDs. The project is
currently evaluating candidate sensors from multiple vendors. 
One of the sensor candidates is a CCD E2V-250, a 4096 $\times$ 4096 pixel 
array that is \SI{100}{\micro\metre} thick and equipped with sixteen
amplifiers. Each pixel is \SI{10}{\micro\metre} on a side and 
has a four-phase structure so as to reach a full-well capacity
of ${\mathrm\sim 160,000\ \mathrm{e}^-}$.

In this study, we utilize exposures from the CCD E2V-250 that were obtained 
by the LSST sensor team during December 2012. The test bench setup uses lamps 
 whose output illumination feeds a monochromator. 
The monochromator output is then either diffused by an integrating sphere
to produce a flatfield, or collimated using a pinhole to produce a spot
(hereafter called ''spot image'').

\section{Broadening of spots and stars with increasing fluxes: The brighter-fatter effect}
\label{sec:bop}

This section presents the amplitude of the broadening of spot or star images
that affects both deep-depleted CCDs and thinned devices.

\subsection{The broadening of spots with increasing fluxes}

The broadening of localized charge distribution when flux increases
can be shown directly from a series of artificial spot exposures.
In figure \ref{fig:spots-diff}, we propose direct evidence for the
broadening of spots with flux on the CCD E2V-250. 
We subtract an average of 20-s
exposure (faint) laboratory spots from an average of 200-s (bright) spot after
scaling the flux to take the difference between exposure times into account.  
The stronger wings and weaker core
of the 200-s spot are clearly visible in the result of the subtraction (figure \ref{fig:spots-diff}, right).

\begin{figure}
\centering
\subfigure[200-s exposures ~~~  (b) 20-s exposures  ~~~ (c) subtraction (a)-(b)]{\includegraphics[width=\linewidth]{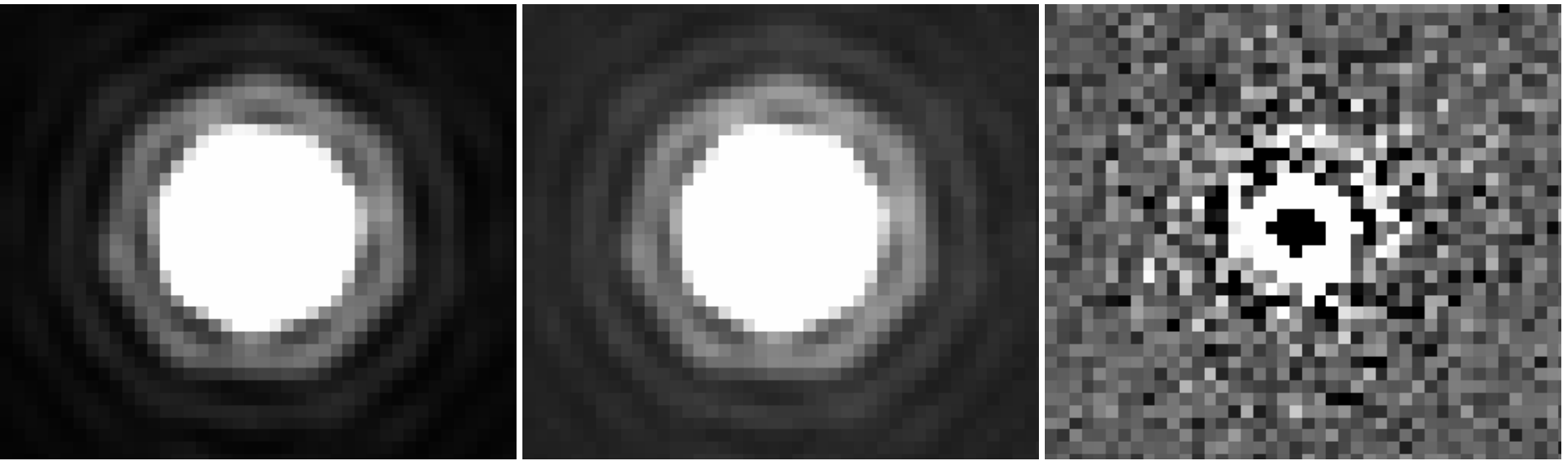}}
\caption{Direct shape comparison of bright and faint spots from the CCD E2V-250. 
The leftmost
image is the average of 200-s exposures; and the middle one averages 20-s exposures;
the rightmost image is the difference after proper scaling  of 200-s spot image 
minus 20-s spot images. The broader wings and lower
peak of the bright spot clearly show up. Since the images required a
small alignment prior to subtraction, we have shifted both averages
by half their separation, so that resampling affects both in the same way.
\label{fig:spots-diff}
}
\end{figure}

\subsection{Measurement of the apparent size of stars}

To perform a quantitative study of the effect, an estimator of size is needed.
 The apparent size of stars or spots is estimated using elliptical Gaussian
weights, whose size are matched to the object \citep[see][\S 3]{Astier13}. The method
is similar, if not identical, to the one proposed in \cite{BernsteinJarvis02}.
Our method turns out to be equivalent to fitting an elliptical
Gaussian to the image with uniform weights. 
We have checked on simulations of non-Gaussian spots of identical shapes with varying S/N ratios, which shows this 
second moment estimator is independent on average of S/N.
The combination of the second moments $M_{xx},M_{yy},M_{xy}$ of the 2-D Gaussian distribution 
defines the image quality (IQ): 
\begin{equation}
IQ = \sqrt[4]{M_{xx} M_{yy}-M_{xy}^2} \notag 
\end{equation}
which we use as to estimate the apparent size of stars.
It should be noted that $|M_{xy}|<< M_{xx}$, $M_{yy}$ and that we also 
refer to the Gaussian root mean square (RMS) $\sqrt{M_{xx}}$ or 
$\sqrt{M_{yy}}$ as the width of a star or a spot for all data sets considered here.

In astronomical exposures, a small correlation between flux and size through color
is expected. On a given field, fluxes are usually 
correlated with color, meanwhile blue stars tend to be broader than red stars 
(because image quality tend to improves in redder wavelengths). This dependency is taken into account
by fitting size versus color relation using low flux stars. The correction is small
on MegaCam data and negligible on DECam.

\subsection{Results}
\label{Results}

Broadening of spots and stars that is seen on all the images and taken with the three
instruments are summarized on the panels of figure \ref{fig:mbf}.
The amplitude of the effect is normalized by the reference sizes
of the spots/stars as a means to compare its impact on the various data samples relative to their IQ.
For each instrument, the intervals that are presented have 
been selected for a dynamic range that is below blooming effects, a degradation 
of the charge transfer efficiency of CCDs that occurs at a level close to full
well, but that varies with CCD readout parameters (see \S\ref{sec:ppc}). 
This directly affects the spot size measurement, with a threshold 
that is around 130 ke$^-$ and 170 ke$^-$,
respectively, for CCD E2V-250 and DECam. This threshold does not show up
on MegaCam data because the ADCs saturate at a lower signal level.

\begin{figure*}
\begin{center}
  \subfigure[LSST - E2V 250 - Spots 550 nm]{\includegraphics[width=0.48\linewidth]{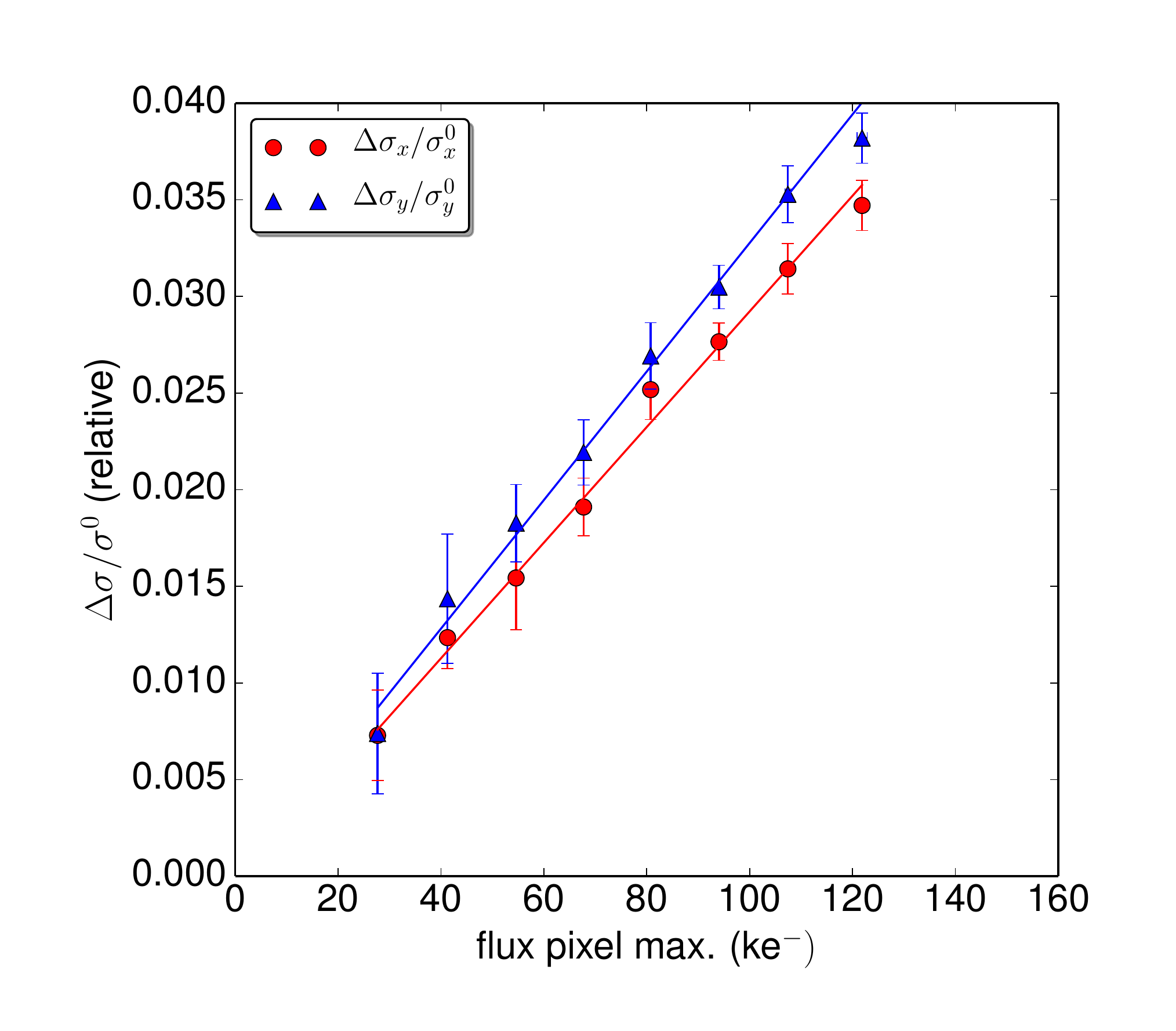}\label{fig:l1}}\quad
  \subfigure[LSST - E2V 250 - Spots 900 nm]{\includegraphics[width=0.48\linewidth]{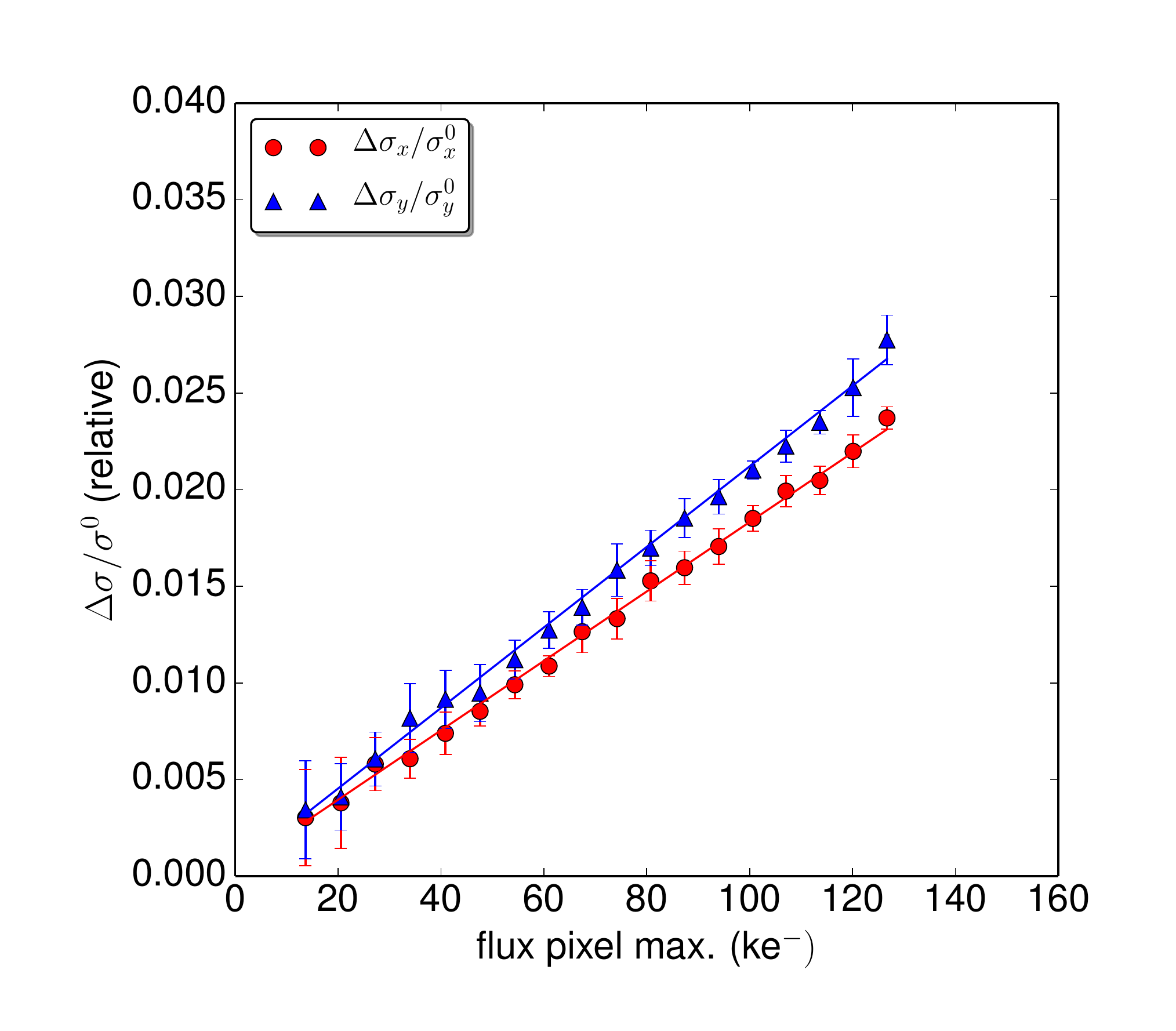}\label{fig:l2}}
  \subfigure[MegaCam - E2V 42-90 - $r$-band stars]{\includegraphics[width=0.48\linewidth]{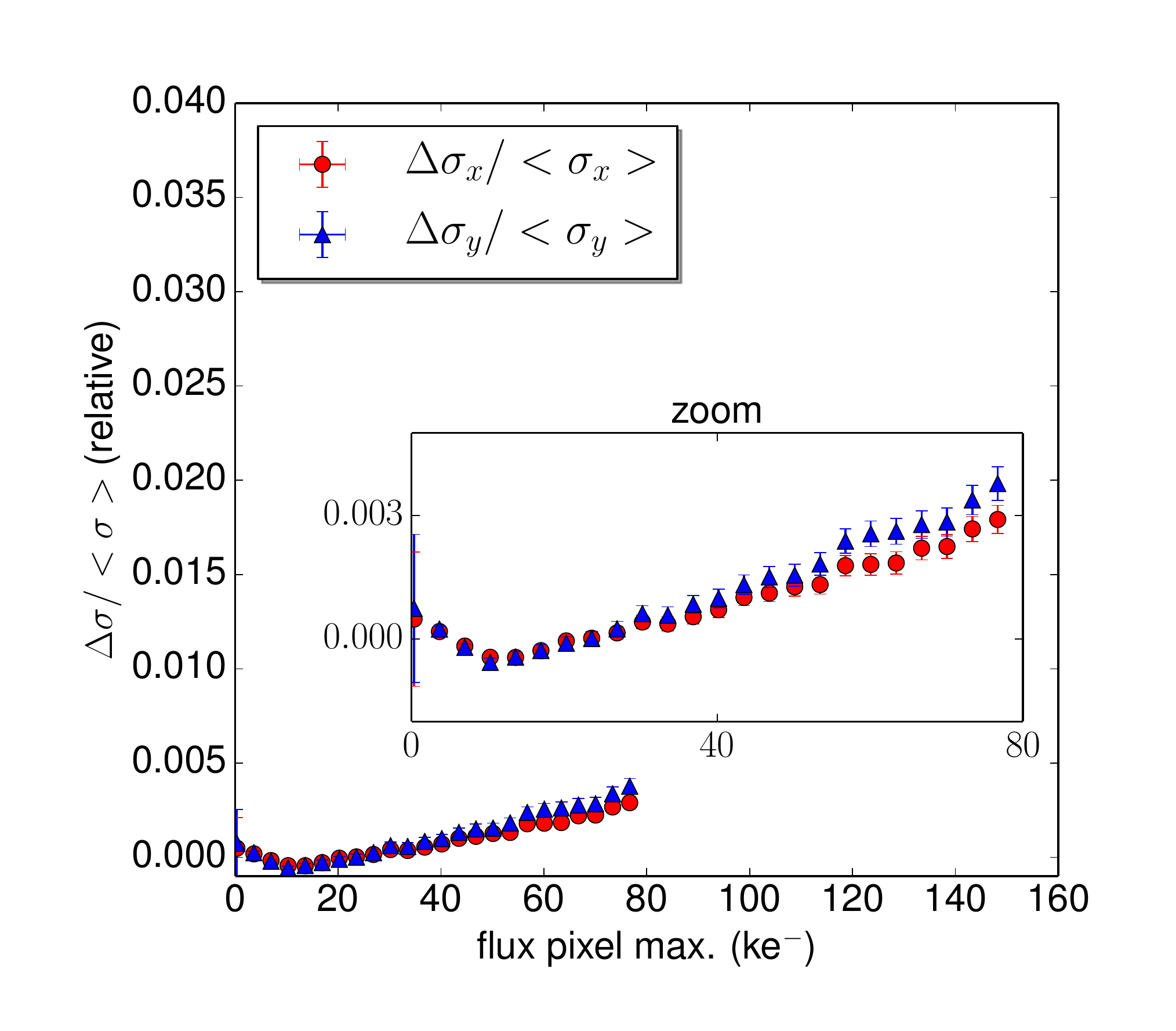}\label{fig:M}}\quad
  \subfigure[DECam - LBL/DALSA - $r$-band stars ]{\includegraphics[width=0.48\linewidth]{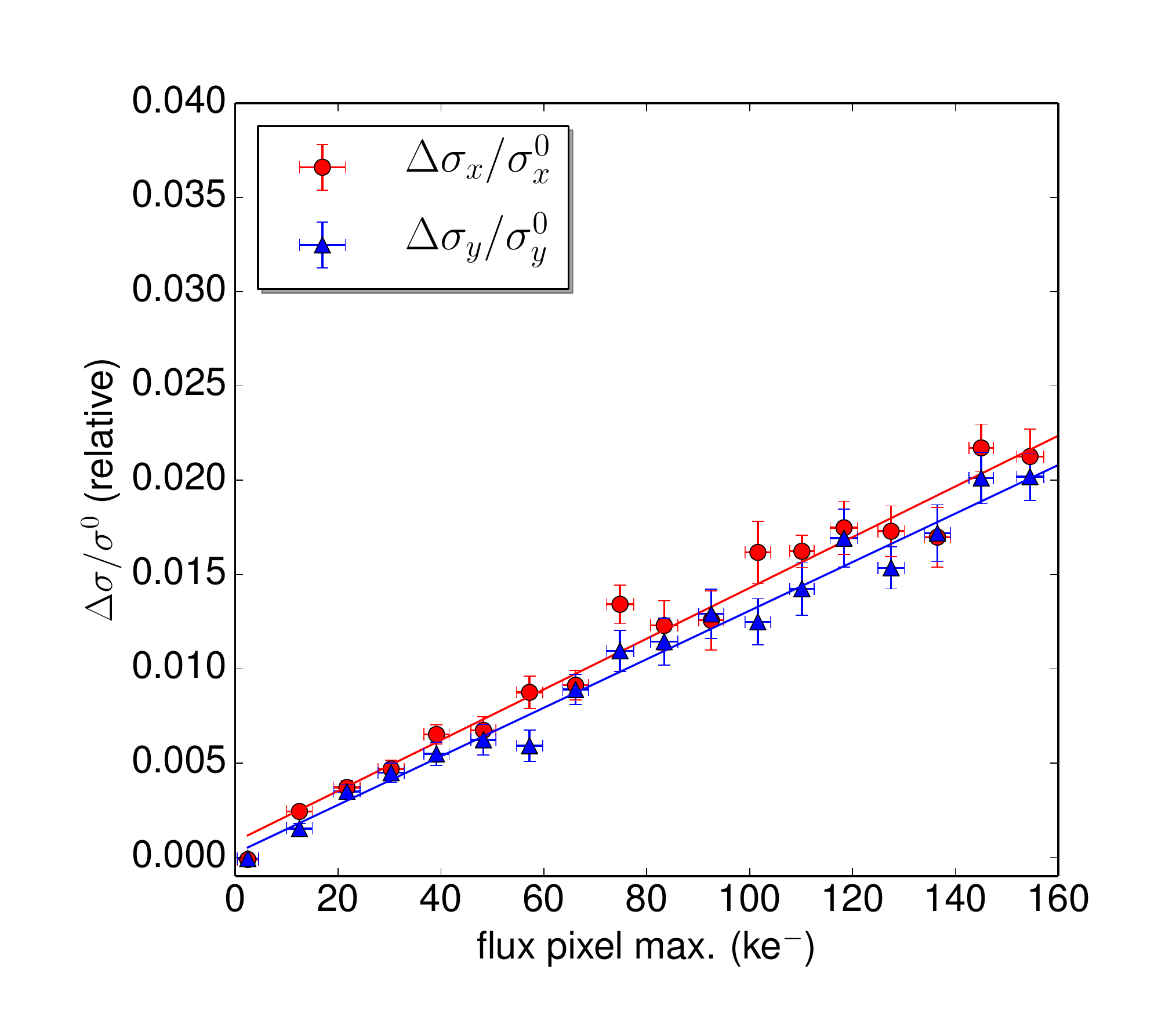}\label{fig:D}}
  \caption{Top: Relative variation of spot width as a function of spot peak flux on CCD E2V-250 at two wavelengths, 550 nm (left panel) and 900 nm (right panel). Bottom left: MegaCam difference of second moments of stars to the average as a function of peak flux. Bottom right: Same measurement on DECam CCD (S11). A small color correction is applied in both cases.  One notes that in all cases the size increases linearly with the peak flux.
 \label{fig:mbf}}
\end{center}
\end{figure*}

The top panels of figure \ref{fig:mbf} show measurements performed on
CCD E2V-250 with 550 nm (left) and 900 nm
``star like'' spots (right).
In both cases, a linear increase of spot size with flux is
visible. From a
vanishing spot to saturation, it amounts to about 3.5\%
for the 550 nm spots and 2.5\% for the 900 nm spots. 
The red spot reference size (which we define as 
the intercept of a linear
fit) is $\approx$ 2 pixels, while the blue spot is $\approx$ 1.6 pixel.
This difference is due to diffraction in the illumination setup. 
Since the spot sizes are different, 
we do not draw a conclusion on any wavelength dependence from these two data sets.

The MegaCam images exhibit an apparent increase of stellar PSF by about 0.008 pixel over
the whole brightness range, which is approximately 0.5\% for this
sample (figure \ref{fig:mbf}, bottom left). It should be noted that this range 
corresponds to half-filled CCDs \citep{doi:10.1117/12.395493}, due to the gain setting of the camera read out electronics.
In this $r$-band on-sky data, we find that the increase rates
are similar along rows and columns with a mild indication that it
might be slightly larger (10 to 20 \%) along columns. 
This plot takes into account the relation that exists between the stars apparent size and color.
The size-color relation is estimated using a set of low flux stars, 
in practice, where a small bin contains a large number of stars. 
The relation between the second moments, and the $g-i$ colors is fitted 
by a linear relation. This polynomial is then evaluated for each
star and substracted to its second moment. This method to subtract the size-color relation 
is replicated from
\citet[\S 10]{Astier13} .

The bottom right panel of figure \ref{fig:mbf} illustrates the effect 
as seen on DECam images using $r$-band images 
on CCD N17. Average PSF here is $\sigma_{x} \approx 1.7$ pixel 
and $\sigma_{y} \approx
1.9$ pixel. It exhibits a broadening of $\approx$ 2\% from zero flux up
to full well, which is slightly lower than what is observed on the CCD E2V-250.
 The size-color relation is taken into account, but its contribution is negligible. 
Over the chips of the focal plane, the average and spread of the 
brighter-fatter effect is (0.0250 $\pm
$ 0.0040) [pix / 100 ke$^-$] in the X (serial) direction and (0.0252 $\pm
$ 0.0037) [pix / 100 ke$^-$] in the Y (parallel) direction. The precision 
of this measurement for a single amplifier is $\approx$ 0.0010 [pix / 100 ke$^-$] in both directions, 
which is at least three times lower than
the spread of the distribution for the entire camera: this may indicate small individual
differences among sensors of the same model or small differences in
operating voltages from sensor to sensor. This could also come from 
inaccurate gain estimations. 

The comparison of the absolue amplitude ot the brighter-fatter effect found on the different sensors 
is summarized in table~\ref{tab:xy_ccd_comp}. The projections in the X and Y direction are indicated in the 
first and second columns, respectively. For the CCD E2V-250, the slopes are found to be larger in the Y direction: by 15\% 
at 900 nm and by 10\% at 550 nm. The observation is more significant at 
900 nm (10 $\sigma$) than at 550 nm (2 $\sigma$) due to a higher statistics. 
For DECam and MegaCam, the effect is also observed to be slightly bigger in the Y direction.

\begin{table*}
\caption{\label{tab:xy_ccd_comp} Comparison of the brighter-fatter effect in the X and Y direction that are observed 
on CCD E2V-250, DECam and MegaCam.}
\begin{center}
\begin{tabular}{c|cl|cl}
\hline
\hline
\multicolumn{1}{c|}{}  &\multicolumn{2}{c|}{X}                   & \multicolumn{2}{c}{Y} \\
Instrument - wavelength      & $\sigma$  &$\Delta\sigma$@100ke$^-$     &$\sigma$ &$\Delta\sigma$@100ke$^-$  \\
                      & (pix)   & (pix)                      & (pix)  & (pix)          \\
\hline
\hline
CCD E2V-250 - 550nm               &  1.594  &  0.047 $\pm$ 0.002      &   1.622  &  0.052 $\pm$ 0.003   \\
CCD E2V-250 - 900nm               &  2.042  &  0.037 $\pm$ 0.0005   &   2.048  &  0.043 $\pm$ 0.0007  \\
DECam - $r$-band ($\sim$ 640nm)   &  1.709  &  0.022 $\pm$ 0.001     &    1.944 &  0.024 $\pm$ 0.001  \\
MegaCam - $r$-band  ($\sim$ 640nm) &  1.980 &  0.005    &    1.960 &  0.006   \\
\hline
\hline
\end{tabular}
\tablefoot{Comparison between the two columns shows that the amplitude of 
the broadening is steeper in the Y direction
than in the X direction. It should be noted that the values that are estimated for the MegaCam instrument are arguable because they depend on the range where they are adjusted.
 }
\end{center}
\end{table*}

Figure \ref{fig:iqv} gathers relative size changes from the whole DECam
mosaic dispersed as a function of IQ. 
It shows the maximum amplitude of the variation in the ($g$, $r$, $i$, $z$, $Y$) DECam filters. 
On the top panel, it can be noted that the
relative amplitude of the effect becomes steeper when the IQ improves. On the bottom panel, 
data points from the various passbands are represented without normalization.
It shows that the absolute amplitude of the broadening is actually 
quite independent of the IQ. It also indicates that the effect is 
mostly achromatic. 
Since it is also reported in \cite{Astier13}
that brighter-fatter slopes are consistent across bands for MegaCam sensors,
it supports the observation that the effect is largely wavelength independent.

\begin{figure}
\centering
\includegraphics[width=1.0\linewidth]{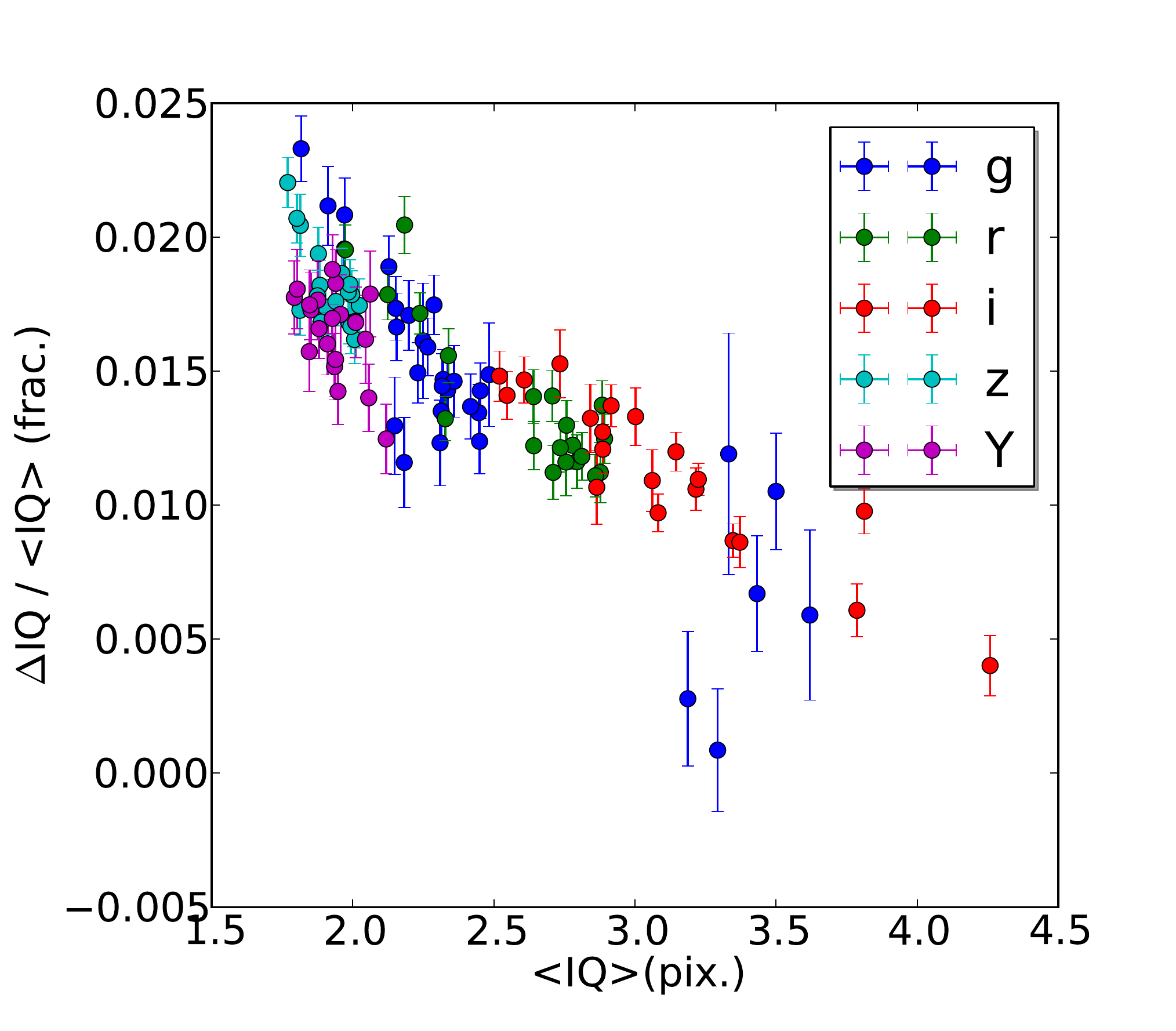}
\includegraphics[width=1.0\linewidth]{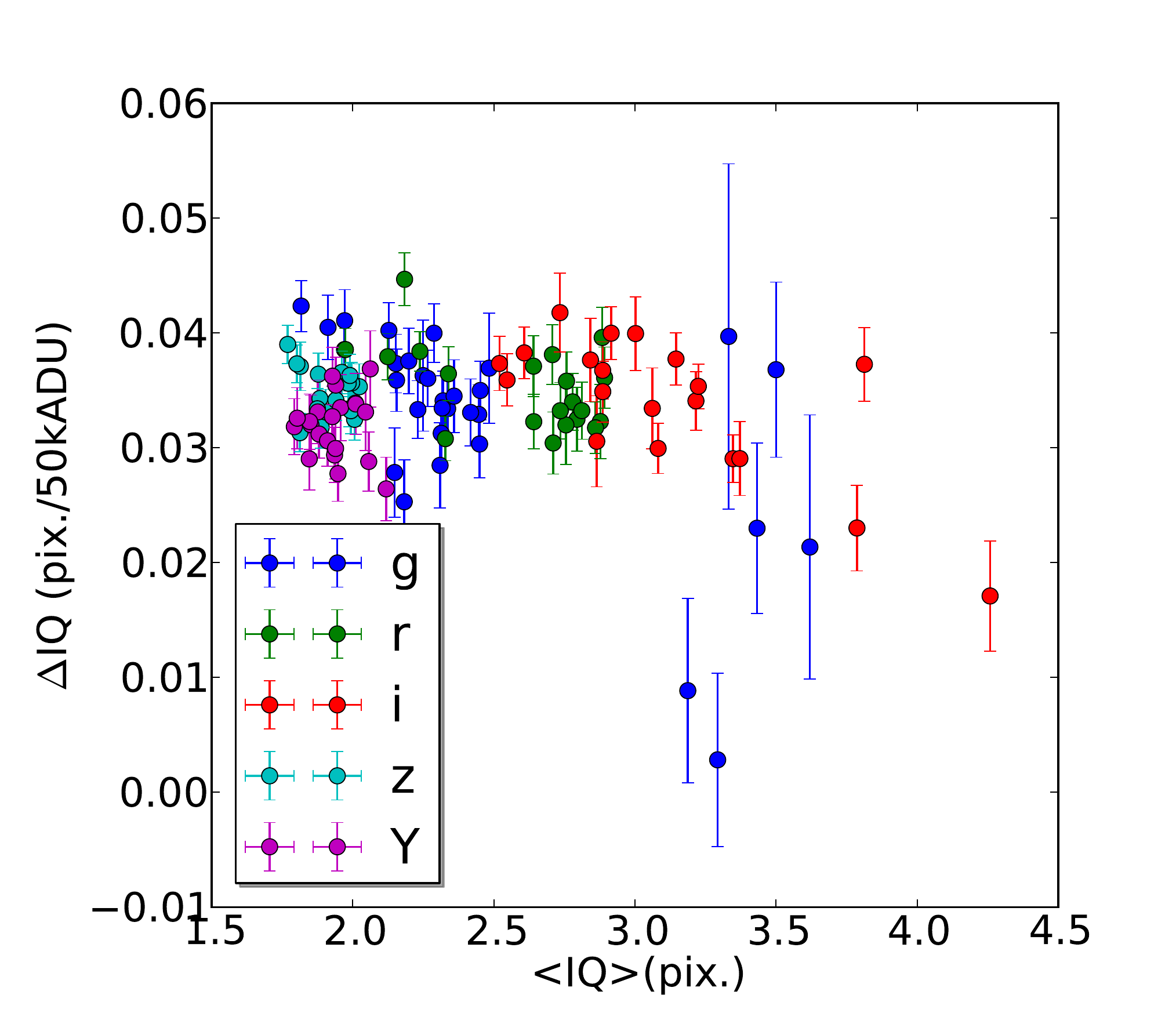}
\caption{Top: Maximum (50 kADU) relative amplitude of the brighter-fatter effect
in DECam. 
The relative slope becomes steeper when the IQ improves.
Bottom: Maximum amplitude of the variation 
of PSF shape in the various filters. The brighter-fatter effect seems 
achromatic, and the linear IQ increase seems fairly independant of IQ.
 \label{fig:iqv}}
\end{figure}

To summarize, the brighter-fatter effect has 
been detected on all the CCDs that we have analyzed.
Even though it is a small effect (few percent of PSFs mean width), it is quite
easy to measure its amplitude with a few percent uncertainty. At this level of precision,
no chromaticity dependence is seen, and a slight anisotropy in X and Y direction 
is observed.

\section{Pixel spatial correlation in flatfield images}
\label{sec:ppc}

The broadening of spots with increasing flux presented in the previous section 
can be depicted as a reduction of the image contrast. 
We see in this section that a similar contrast reduction also appears in flatfield 
images. It manifests itself as a non-linearity of the photon
transfer curve (PTC). A PTC is the representation of the variances versus the average fluxes 
of flatfield illuminations obtained with increasing exposure times. Considering only 
Poisson noise, the relation between the two observables is expected to be linear.  
However, we see that a significative departure from linearity is actually observed 
and that it is associated with 
linearly increasing pixel correlations. 

\subsection{Non-linearity of the photon
transfer curve (PTC)}

The non-linearity of the photon
transfer curve is illustrated in figure \ref{fig:ptc} using 
eight different segments of the CCD E2V-250. The raw PTCs (on left panel) can be rescaled 
(on right panel) using a classic method with linear fit of the PTC at low flux level 
to determine the read out gain from the relation:
\begin{equation}
G = \mathrm{flux}/\mathrm{Variance},
 \label{eq:gain}
\end{equation}
where $G$ is expressed in $(e^{-}/ADU)$. 
When PTCs are rescaled to the same slope at the origin, 
the non-linearities clearly appear and are found to have similar
magnitude in all channels (figure \ref{fig:ptc}, right panel). 
When the residuals of the PTCs to the Poissonian noise are represented (figure \ref{fig:ptc-res}),
it is also seen that the phenomenon does not occur at a given threshold, 
but rather sets on from the very beginning of the PTC.
 
The non-linearity of the PTC is an unexpected feature: flatfield images are naively expected to exhibit
a Poissonian noise, which is a variance scaling with the average.
However, the variance measured 
at high-flux is actually significantly lower 
than expected from extrapolating the variance
of low-flux flatfields according to Poisson law: the discrepancy is as high as 20\% 
in the case of the CCD E2V-250. This effect has long been observed
on other CCDs \citep{Downing06}, but we have not been able to 
find a physical explanation in the literature. 

\begin{figure*}
\begin{center}
  \subfigure[PTC in ADU]{\includegraphics[width=0.48\linewidth]{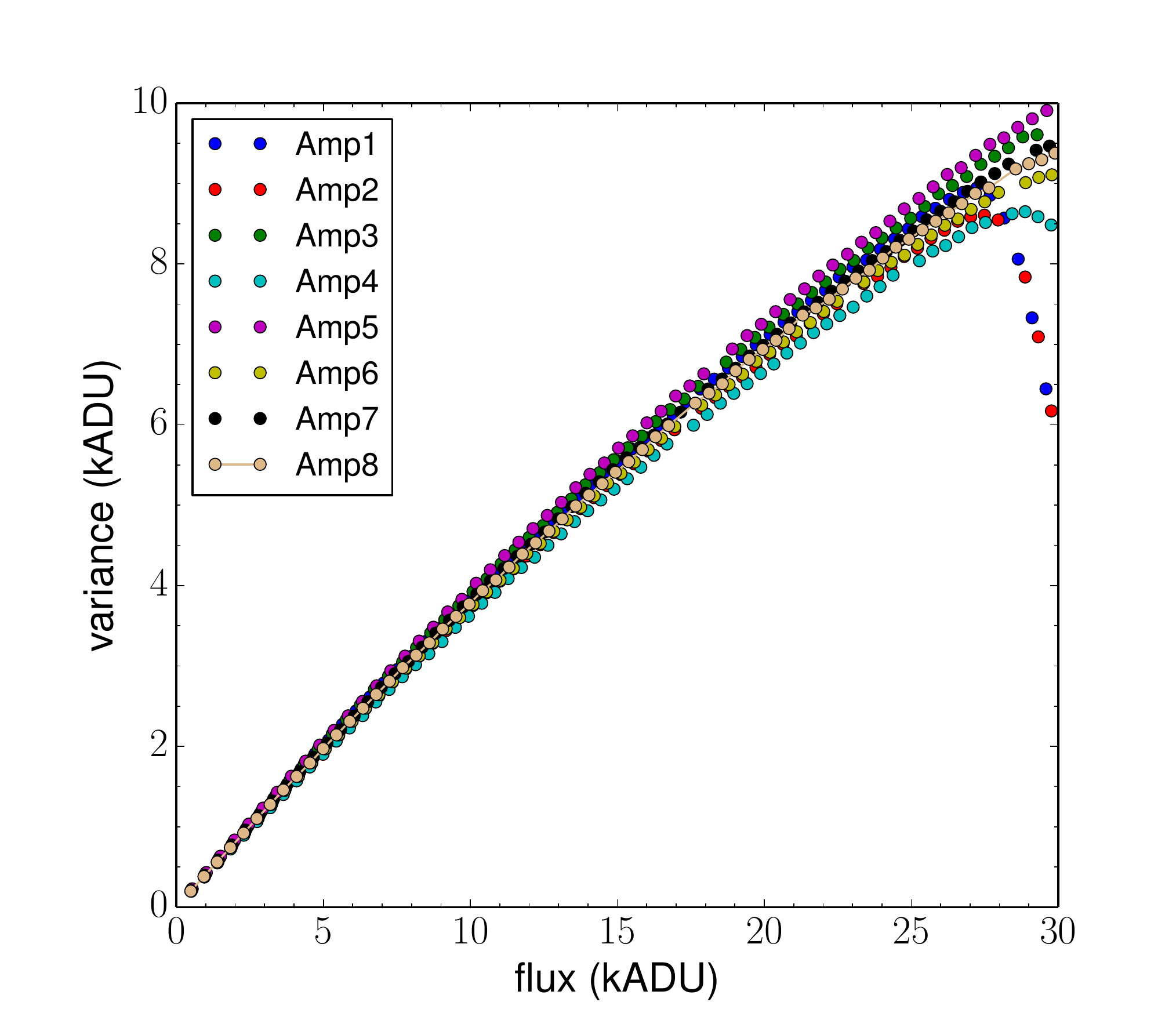}\label{fig:ptc1}}\quad
 \subfigure[PTC in e$^-$]{\includegraphics[width=0.48\linewidth]{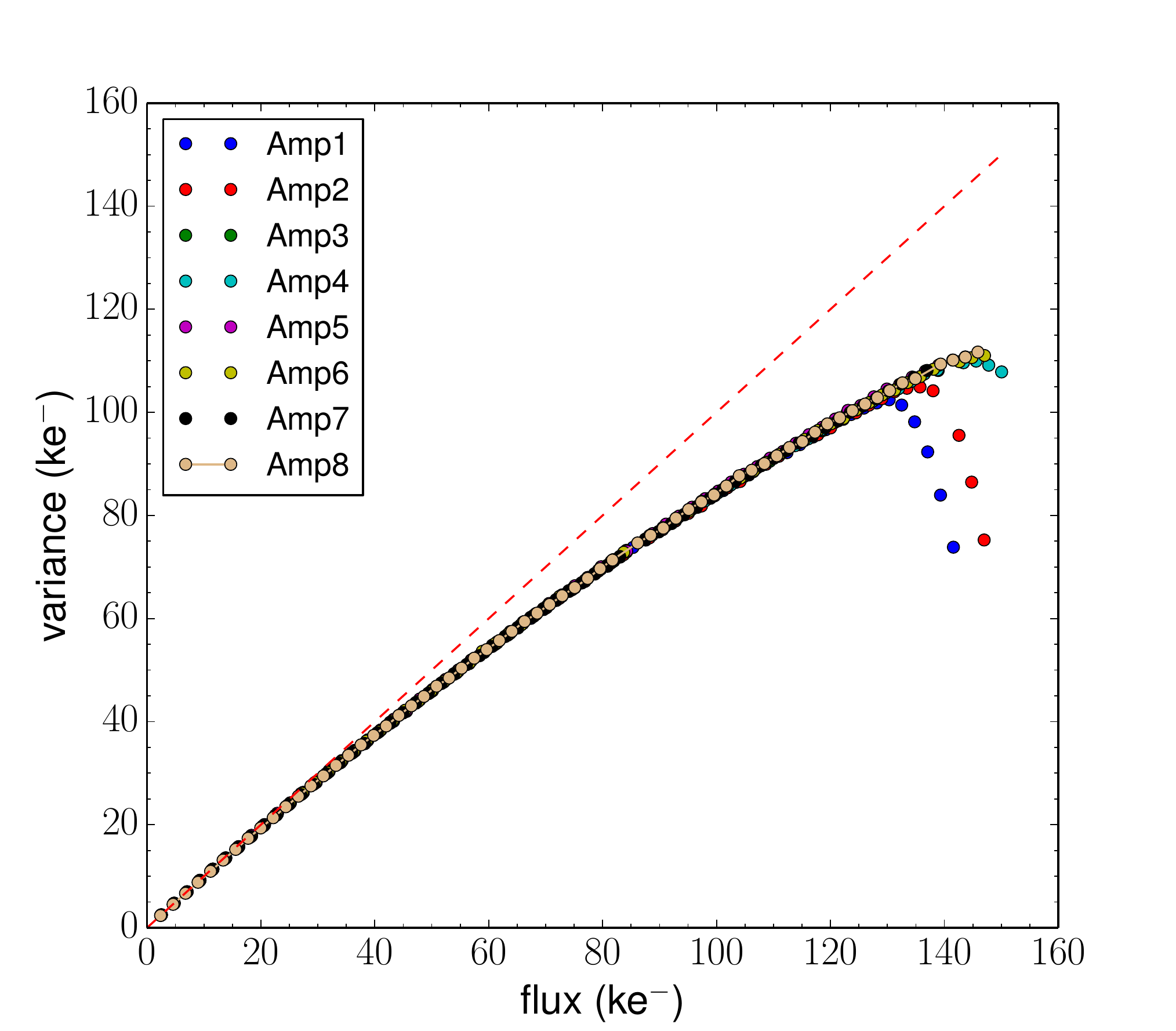}\label{fig:ptc2}}\quad
\caption{(a) Measured photon transfer curve of amplifiers \#1 to \#8 
of the CCD E2V-250 expressed in ADU. (b) PTCs are converted in e$^-$ using the readout gain
measured as the inverse of the slope at the origin (see \ref{linearly}).
Because departure from linearity of the PTCs is similar 
for different amplifiers this indicates
that the cause of the effect is to be found in the CCD itself rather 
than in the electronic readout.}
  \label{fig:ptc}
\end{center}
\end{figure*}

\begin{figure} 
  \includegraphics[width=0.96\linewidth]{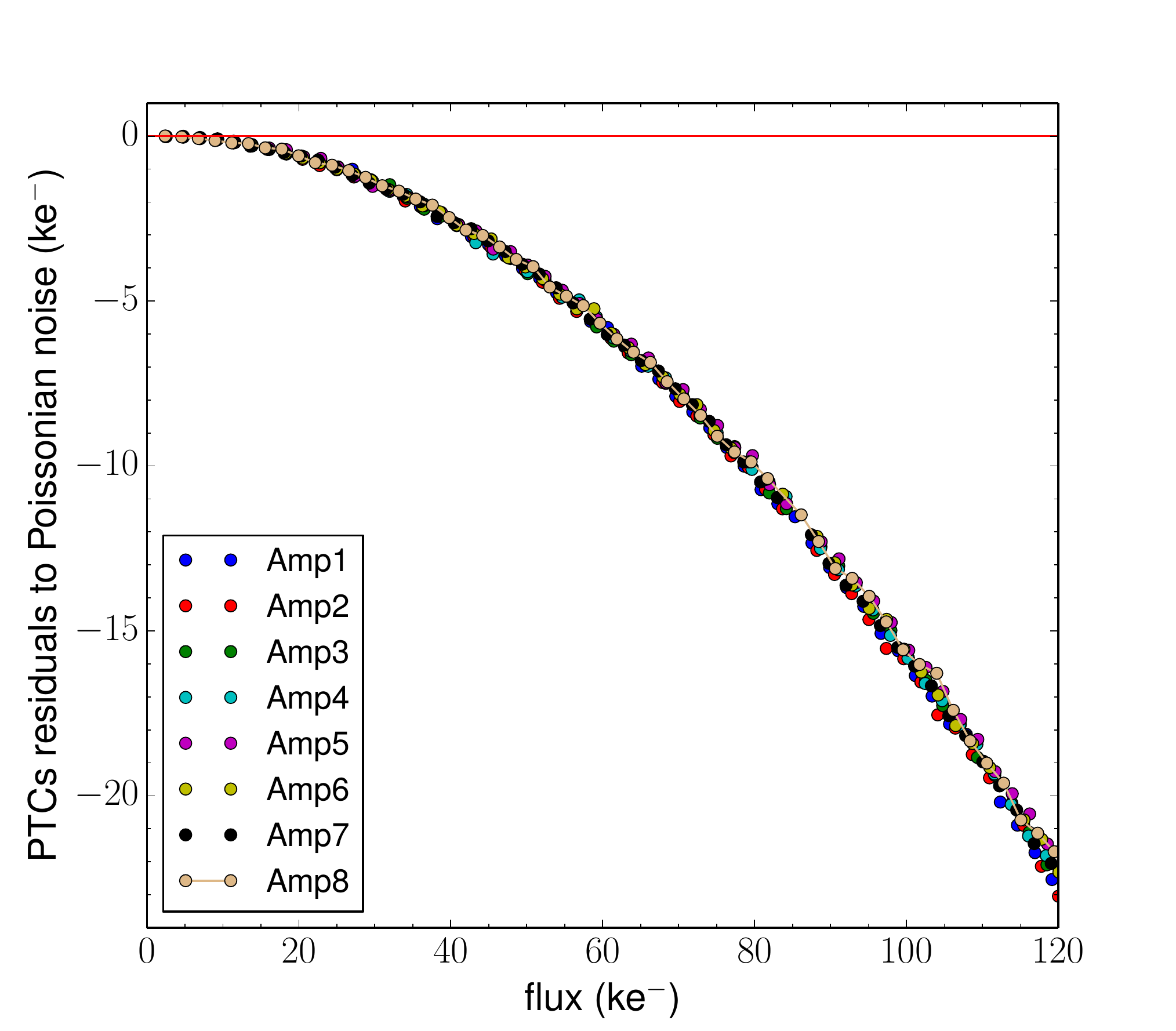}
\caption{ CCD E2V-250: Residuals of the PTCs to the expected Poissonnian noise.
  The departure from linearity starts from the very beginning.}
  \label{fig:ptc-res}
\end{figure}

\subsection{Measuring pixel spatial correlations}

\subsubsection{Pair images subtraction} Correlation coefficients are computed
from the difference of two flatfield images that have received the same overall illumination. 
This is a classic technique to remove apparent
correlations due to non-uniformity of the flatfield image that is typically
due to pixel size variations, QE variations, or spatial variations of
the illumination itself. Temporal stability of the light source at the
 per mil level is then mandatory. 
In this analysis, this is why flatfields from artificial sources are being used rather 
than twilight flatfields.

\subsubsection{Statistical precision} Spatial correlations between pixels 
are evaluated using covariances
normalized by variance. $R_{k,l}$
refers to the correlation coefficient between pixels separated by $k$
columns and $l$ rows. It is important to emphasize that $k$ and $l$ do not 
refer to independent variables but index the spatial relation between two pixels,
so $R_{k,l} \neq R_{l,k}$. The statistical precision on 
any given correlation coefficient is $1/\sqrt{N}$,
where $N$ is the number of pixels. The statistics is doubled for the off-axis
correlations ($k,l \neq 0$) by combining the measurements of two quadrants ($R_{k,l}$ and $R_{k,-l}$
for instance).
On MegaCam and DECam, each 
amplifier channel has 4 MPix while CCD E2V-250 reads 1 MPix per amplifier channel.
This gives uncertainties of 0.5$\permil$ and 1 $\permil$ respectively.
Statistical precision is further increased by using many pairs of flatfields. 
The PTC from the CCD E2V-250 
contains $\approx$ 100 points, which allows us to improve 
precision on correlation measurement down to $1\times 10^{-4}$.
A smaller improvement is obtained in the case of DECam and MegaCam, where PTCs were acquired 
in situ, and gather only $\approx$ 20 points. The precision on correlation measurement
on MegaCam is also reduced due to a partial readout of each channel (only 100 kPix).
Anyhow, measuring pixel spatial correlations requires
some care due to the existence of miscellaneous effects that generate
spurious pixel correlations.

\subsubsection{Spurious pixel correlations} 
Wherever hot or dark columns are being detected, they are masked. 
 Moreover, CCD images sometimes exhibit localized image contrasts 
that are not attributed to the illumination itself but rather
to CCD defects or to specific setups (clocking voltage (CV) 
and backside substrate (BSS) voltage) of the controller used to drive the CCD.
They necessitate an elaborated masking algorithm, since the level may be very
close to mean illumination and the patterns are multiforms. First, we detect the pixels that 
are 3$\sigma$ above a local average \footnote{The average and $\sigma$ of
    pixels are computed over rectangles of 128$\times$128 pixels,
    once using all pixels, and a second time using only pixels within
    2$\sigma$ of the average.}. Then, these pixels are masked if there are more than two other
pixels above 3$\sigma$ in their $\pm$ 1 pixels surrounding. These masks are applied 
on both images from a pair and on its difference.

Correlations that cannot be attributed to any specific 
feature in the image can be otherwise detected from a non-zero intercept when 
fitting a given correlation coefficient versus flux. For instance most channels have 
more than 3$\sigma$ residual
anti-correlation $R_{1,0}$ at zero flux in
our LSST data set, while all the other coefficients have values 
compatible with zero. Considering that 
this offset is only affecting consecutive pixels in the readout sequence
and that it manifests itself even without illumination, 
we assumed that it is introduced by the readout sequence 
(because of an incomplete reset of the baseline, for instance).
Such an effect would generate linearly increasing covariances with flux, 
which translate to a constant correlation contribution that 
is subtracted from all measurements of $R_{1,0}$.

\begin{figure}
\centering
\includegraphics[width=0.96\linewidth]{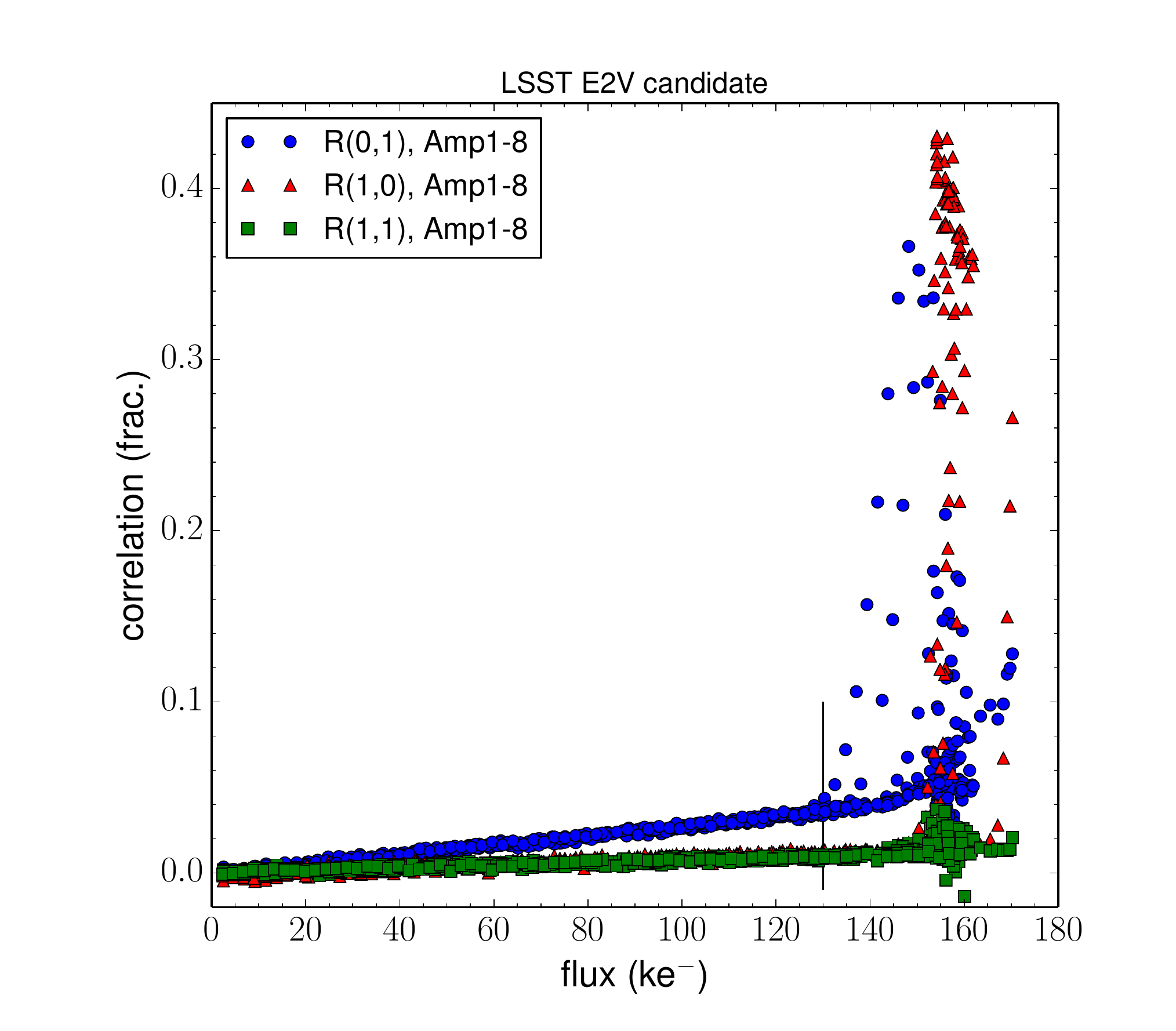}
\caption{Superposition of the evolution with flux for nearest pixel correlations
(coefficients  $R_{0,1}$  $R_{1,1}$  $R_{1,0}$) 
for read out channels 1 to 8 of CCD E2V-250. 
The whole dynamical 
range shows a threshold where correlations strongly increase: First $R_{0,1}$, a correlation in Y 
direction, and near full well, $R_{1,0}$, a correlation in X direction. The diagonal 
correlation $R_{1,1}$ does not exhibit 
any threashold. This is expected since neither transfer 
nor read out could contribute to correlate pixels in this direction.
The vertical dark line indicates the early threshold of $R_{0,1}$.
  \label{fig:chip1a81}}
\end{figure}

\subsubsection{Correlations at the high-flux level} 

On the high-flux end of the dynamic range, 
correlation features are expected to appear when the pixel
contents are approaching full well. These are seen as blooming effects
that increase up to saturation. These correlations have a characteristic
signature: they appear when a given threshold is reached (that varies slightly
from one read out channel to the other), first in Y direction, then in
X direction and in diagonal but with a much smaller amplitude in the latter. 
Figure
\ref{fig:chip1a81} illustrates these three features with the CCD E2V-250: 
$R_{0,1}$ (in red)
coefficient greatly increases at levels between $\sim$ 130 - 160 $ke^{-}$,
depending on the amplifier channel. Near the full well, strong horizontal
correlations $R_{1,0}$ are visible (blue triangles). Lastly, the correlation
$R_{1,1}$ (in green) shows a quite linear behavior up to full well.
Seen from this perspective, the dynamic range is not 
simply divided into two regimes with a low range, where pixels linearly respond to
illumination, and a high range near full well, where pixels saturate.
For this CCD, an earlier threshold occurs around 130 ke$^{-}$, where $R_{0,1}$ abruptly increases.
It also corresponds to the extremum 
of the PTCs (figure \ref{fig:ptc}). In the next section, we focus our 
analysis of pixel spatial correlations
on the dynamic range below this flux level (indicated by the vertical dark line 
on figure~\ref{fig:chip1a81}).

\subsection{Linearly increasing pixel spatial correlations}
\label{linearly}

\subsubsection{Pixel correlation maps} \label{par:pix}
Pixel spatial correlations 
are detected on the three instruments up to a distance of 4 pixels. 
They are presented for CCD E2V-250, 
DECam, and MegaCam on  figure  \ref{fig:2dplot2}. For LSST and DECam, 
the correlation $R_{0,1}$ is about three times larger than 
$R_{1,0}$, while they are of the same order for MegaCam. In all cases, this 
anisotropy between (Y) direction and (X) direction vanishes at larger distances.
At a separation of 4 pixels, all correlations of all CCDs are as low as a few 10$^{-4}$,
which approaches the limit of sensitivity of the measurements. 

For the CCD E2V-250, it has been shown by \cite{BNL13}
that an increase of the parallel clocking voltage (CV) from 8V to 10V decreases the 
level of the correlation $R_{0,1}$ while keeping the other coefficients unchanged.  
In this paper, we complete the study of the impact of varying pixels' electrodes voltage by  
repeating the measurements of the correlations with various backside substrate (BSS) voltages.
The correlation with the next pixel in the parallel direction
$R_{0,1}$ (top panel of figure \ref{fig:bss}) increases as the BSS is decreased down to 10-20 V; 
below this level, the correlation starts decreasing. On the same 
interval, the correlation coefficient with 
next pixel in the serial direction $R_{1,0}$ (middle panel) decreases
and shifts to negative values below 10-20 V. The other correlation coefficients monotonously
increase as BSS decreases (the bottom panel of figure \ref{fig:bss} illustrates this with $R_{0,2}$ and $R_{2,0}$); 
it is worth noting that diffusion mechanisms (as suggested in \cite{ma}) cannot explain 
the evolution with BSS and CV, the amplitude, the anisotropy, and the long range of these correlation coefficients. 
In contrast, the predictions from a simple electrostatic simulation 
are compatible with these observations (see \S \ref{elecsim} and \S \ref{ids}).

\begin{figure*}
\begin{center}
  \subfigure[LSST - CCD E2V 250]{\includegraphics[width=0.31\linewidth]{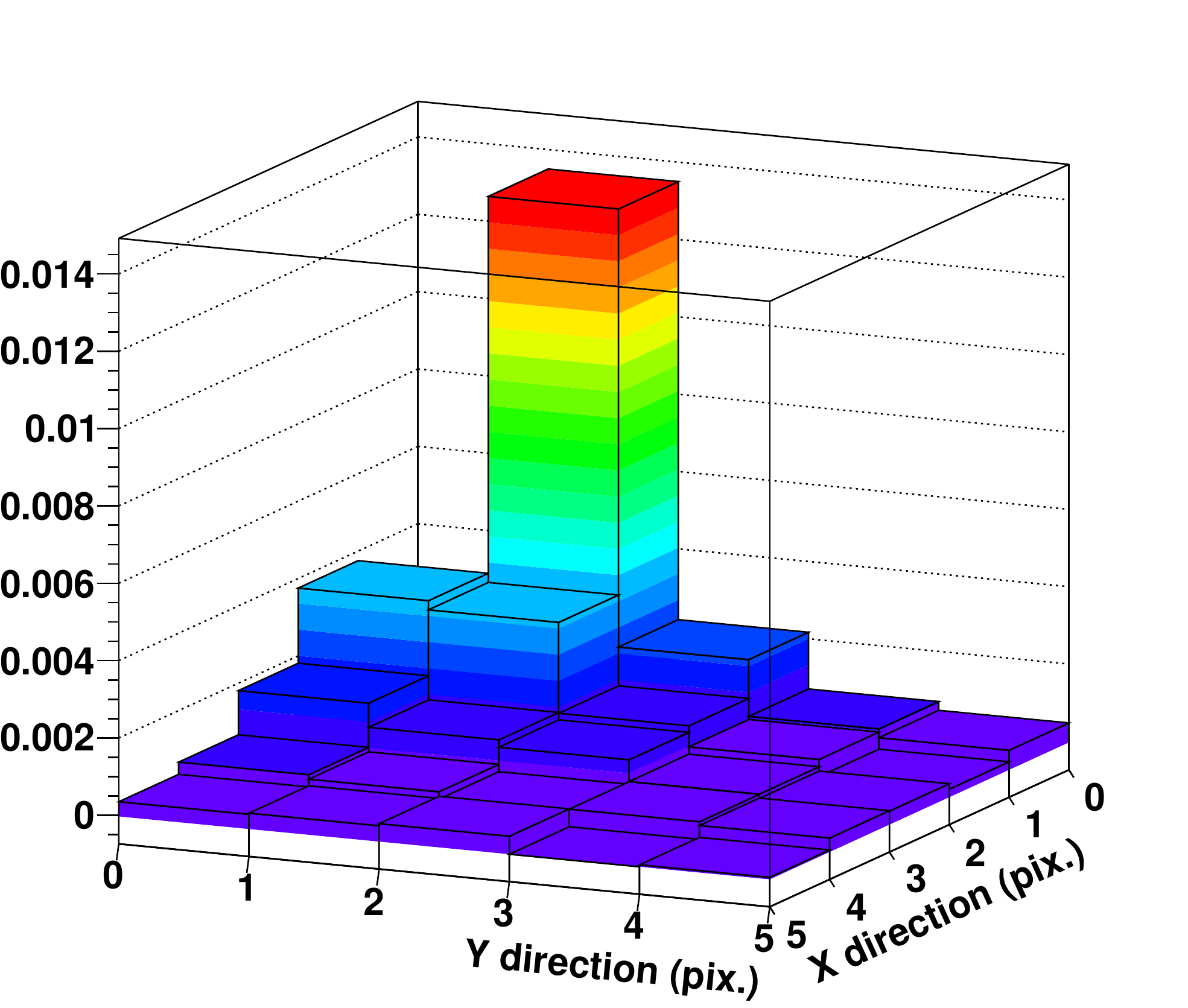}}\quad
  \subfigure[DECam - CCD S11]{\includegraphics[width=0.31\linewidth]{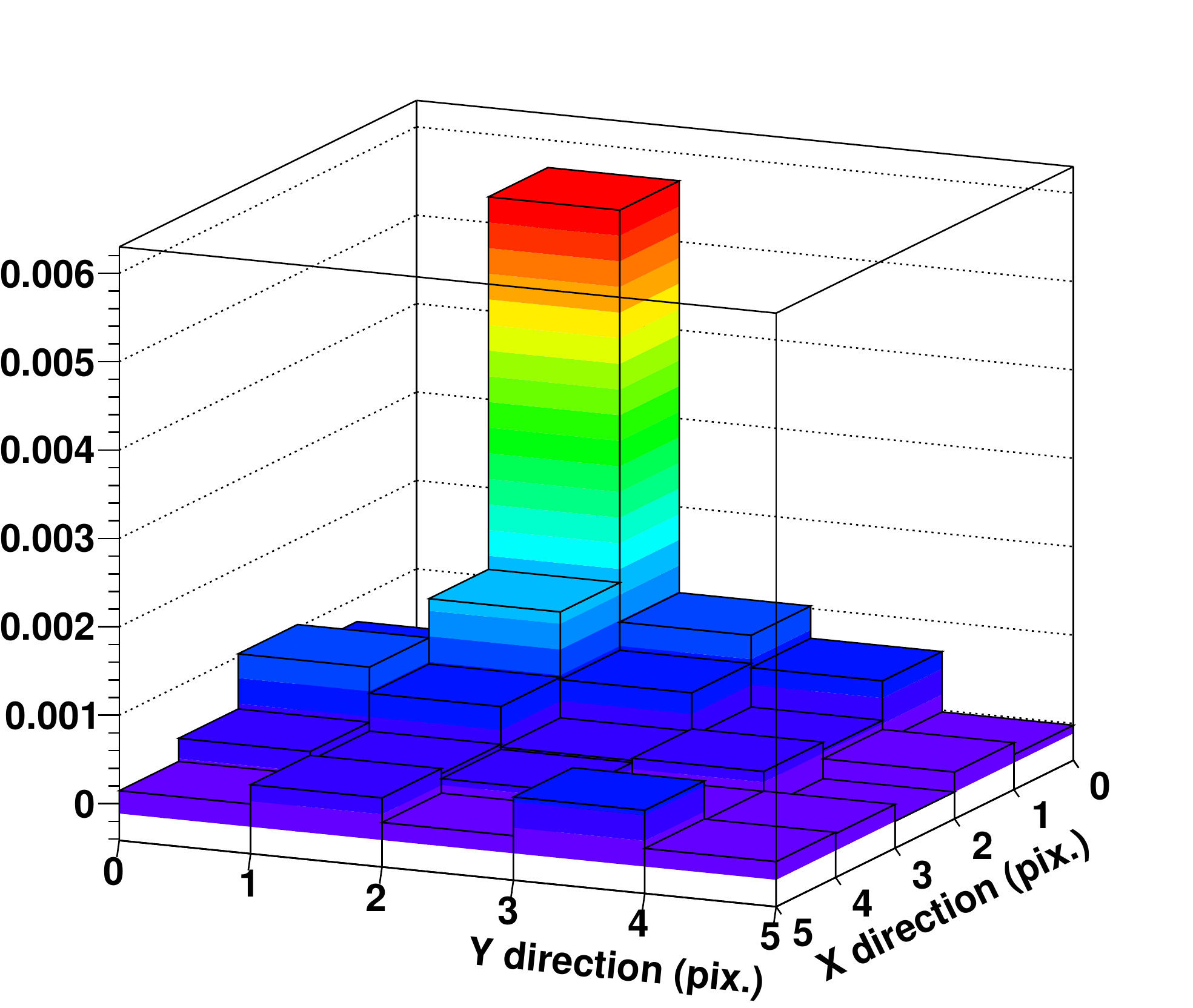}}\quad
  \subfigure[MegaCam]{\includegraphics[width=0.31\linewidth]{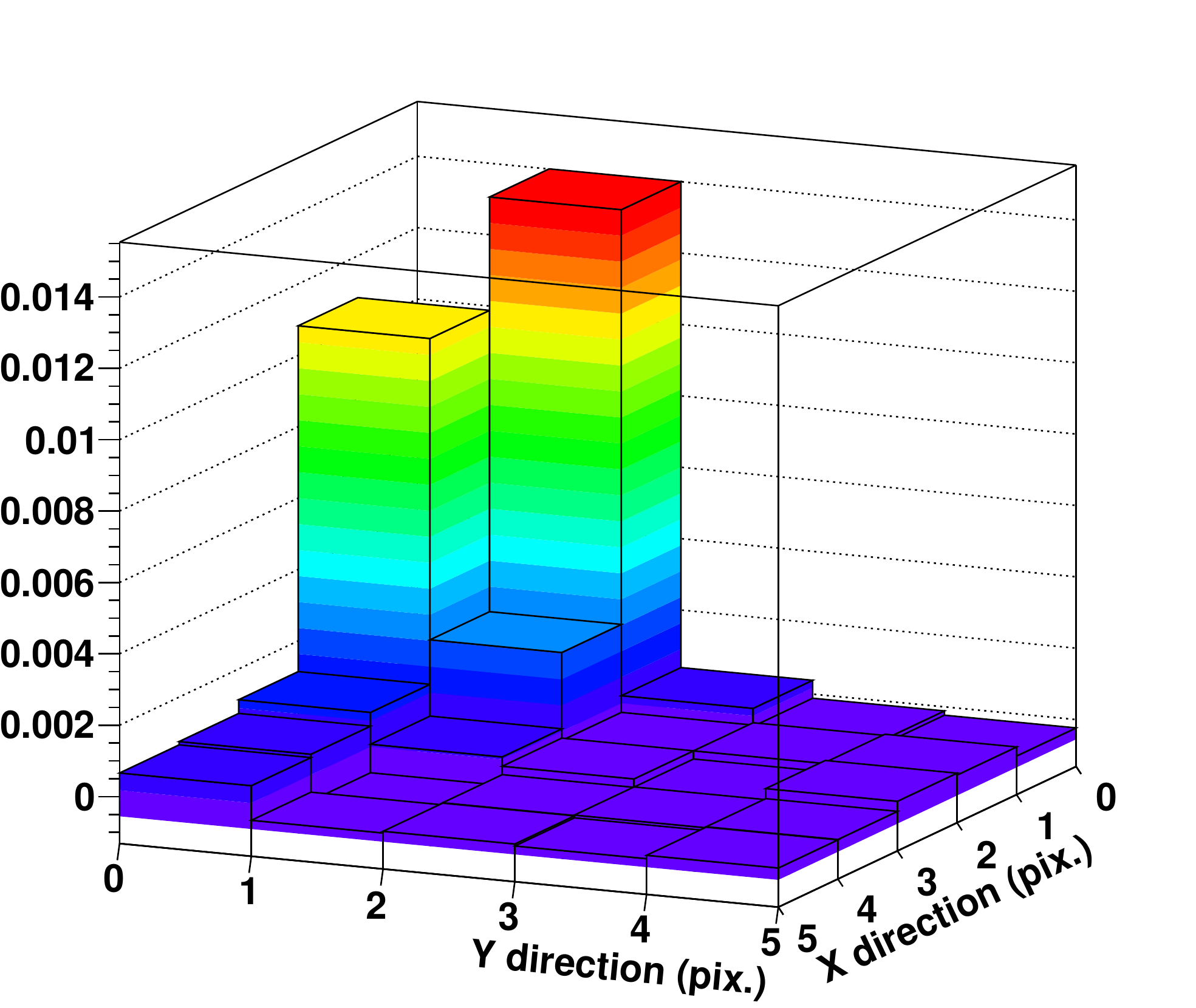}}
\caption{ 2-D correlation map in flatfields at a 50 ke$^-$ level 
for (a) CCD E2V-250 (channel 4, the 
channel on which spots illumination presented in the
first section were done), (b) DECam (CCD S11), and (c) MegaCam. 
Row and column, respectively, correspond to label $X$ and $Y$. An 
anisotropy between the amplitudes of the $R_{0,1}$ and  $R_{1,0}$ coefficients 
is observed. For distance separations larger than 1 pixel, this anisotropy tends to vanish 
($R_{X,Y} \approx R_{Y,X}$ for $X$ and $Y$ > 1).
  \label{fig:2dplot2}}
\end{center}
\end{figure*}

\begin{figure}
\centering
\includegraphics[width=0.96\linewidth]{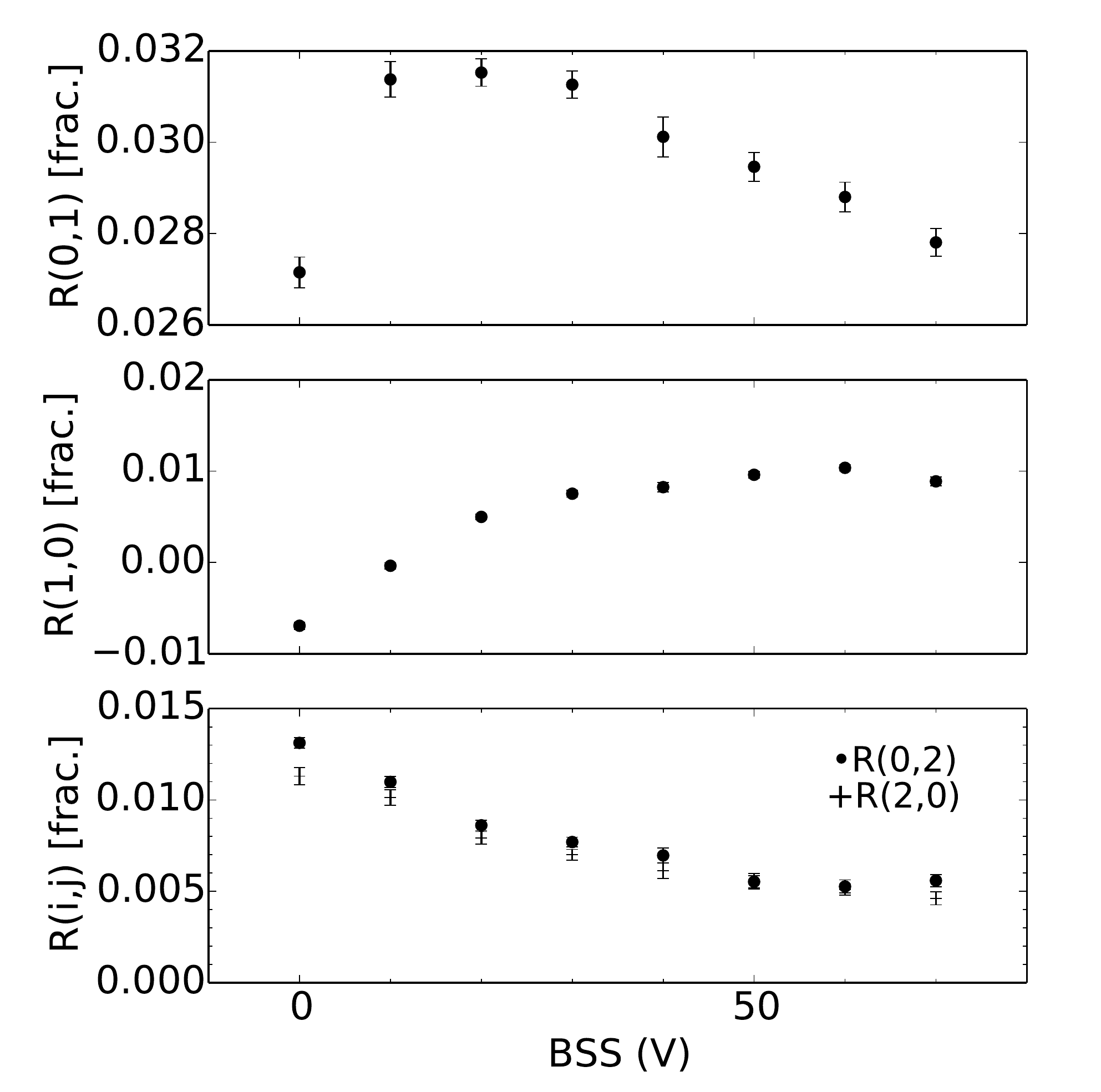}
\caption{Variation of correlation coefficients (measured at 100 ke$^-$)  
with respect to BSS measured on the CCD E2V-250. 
Top panel: $R_{0,1}$. Middle panel: $R_{1,0}$. Bottom panel: $R_{0,2}$ and $R_{2,0}$.
The other long range correlations behave like $R_{0,2}$ and $R_{2,0}$: they decrease
as the BSS voltage increases. 
  \label{fig:bss}}
\end{figure}  

The pixel correlation maps are found to scale with fluxes
for all detectors. This is illustrated in figure
\ref{fig:chip1a82} by zooming on the range 0 to 130 ke$^-$ for 
the eight channels of the CCD E2V-250 that were presented
in figure \ref{fig:chip1a81}. There is no evidence for chromaticity 
dependance of this general trend: a linear fit of $R_{k,l}$ variations for PTC ramps obtained with illumination
 at 500 nm, 700 nm, and 
900 nm wavelengths has shown no evolution with respect to wavelength. 
For instance  $R_{0,1}$ slope is found to be   
  (1.39~$\pm$~0.04)~$\times 10^{-6}$~[frac / ADU] at 500 nm wavelength,
  (1.39~$\pm$~0.04)~$\times 10^{-6}$~[frac / ADU] at 700 nm wavelength, and
  (1.42~$\pm$~0.03)~$\times 10^{-6}$~[frac / ADU] at 900 nm wavelength.
They are all compatible within 1$\sigma$ RMS; likewise,
other correlation coefficients have no detectable variation either.

\begin{figure}
\centering
\includegraphics[width=0.96\linewidth]{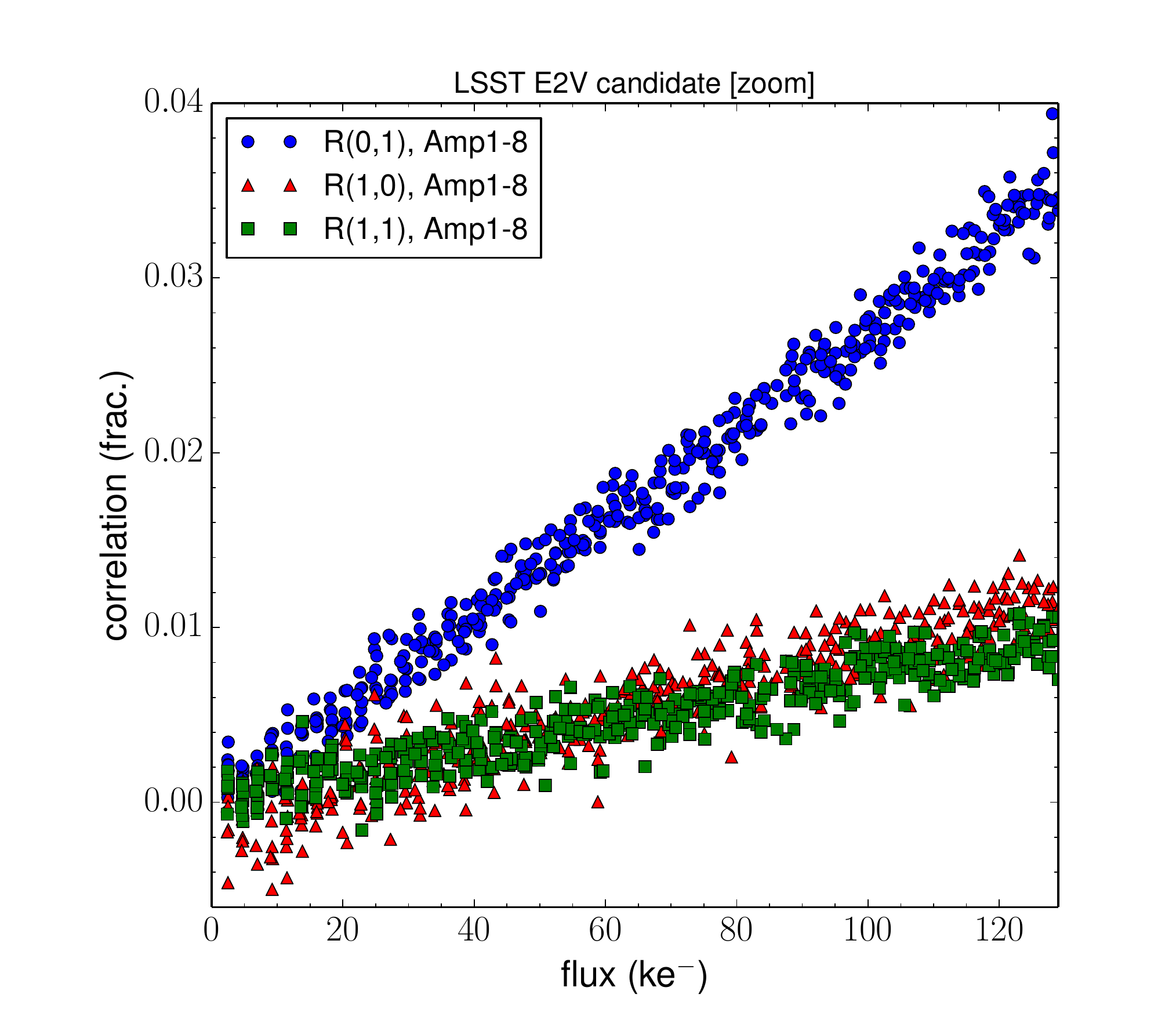}
\caption{Superposition of evolution with respect to flux of coefficients  
$R_{0,1}$  $R_{1,1}$, and $R_{1,0}$ on 
the dynamical range below the threshold, as indicated by the vertical black line 
of figure \ref{fig:chip1a81}. 
On this interval, all  the linearly 
increasing correlations that are discussed in this section shows a monotonic behavior.
On this CCD E2V-250, most channels exhibit a small ($\approx$ -~0.003), but significant 
anti-correlation pedestal in $R_{1,0}$ (Y-intercept), an offset that is not seen 
with the others coefficients nor with the other sensors. We attribute it 
to the electronic chain used to collect this data, and we subtract it to
the actual measurements.
  \label{fig:chip1a82}}
\end{figure}

\subsubsection{Relation between PTC non-linearity and linearly increasing correlation}
\label{ptc}

It has been verified that the process that correlates the pixels 
also conserves charges. This is straightforward
to see from a linear fit of flatfield mean flux versus exposure time. For instance 
the departures for the eight channels of the CCD E2V-250 from linearity of response on the range from 0 to 130 ke$^-$ are below 2$\permil$ 
peak to peak. 

The connection between non-linearity of the PTC and the linearly increasing correlations
can be most clearly illustrated by summing all the correlations and adding them to the 
PTC. In appendix \ref{recovar}, we assume a conservation of charge
to demonstrate that the difference between the raw PTC and the Poisson variance
is the sum of the covariances. This conclusion has a consequence on the measurement 
of the gain that was given by the relation (\ref{eq:gain}). It is then modified in 
the following way: 
\begin{equation}
V_{raw}(N_{ADU}) = - \alpha N^2_{ADU} + \frac{1}{G} N_{ADU} \; . \notag
\end{equation}
At this point, $\alpha$ is an empirical parameter that is introduced to 
describe the quadratic behavior of the PTC, which is expected given a 
linear rise of the correlations. We refer to the end of \S \ref{sec:algebra}
for an interpretation of this parametrization using our model. 
The variable $G$ is obtained by a second degree polynomial 
fit on a range up to the PTC extremum, instead of the linear fit
on an arbitrary low flux interval. It should be pointed out that the 
gain discrepancy 
between both methods can be as large as
10\% depending on what is assumed to be the low flux interval.
For this new method, the relative uncertainty on the gain estimation is 
found to be between 3-4 $\permil$ half of which comes from shot noise, while the rest 
corresponds to the 2$\permil$ departure from linearity of response.

Figure \ref{fig:varprof} shows that the linearity of the PTCs
are restored when the variance and the covariances are added together. The plot 
is expressed in e$^-$, using the gain as measured by the method 
presented above. At a 100 ke$^-$ flux level, the covariances (up to 4 pixels in distance)
add up to 18\% of the variance.
The dashed red line that appears on the plot indicates the expected photon noise.
Its slope is $\approx$ 0.5\% above the combination of variance plus covariance, which
is consistent with a truncation of the sum of small residual correlations 
at distances larger than 4 pixels. It indicates that they correspond
to less than 3\% of the total pixel correlations.

\begin{figure}
\centering
\includegraphics[width=0.96\linewidth]{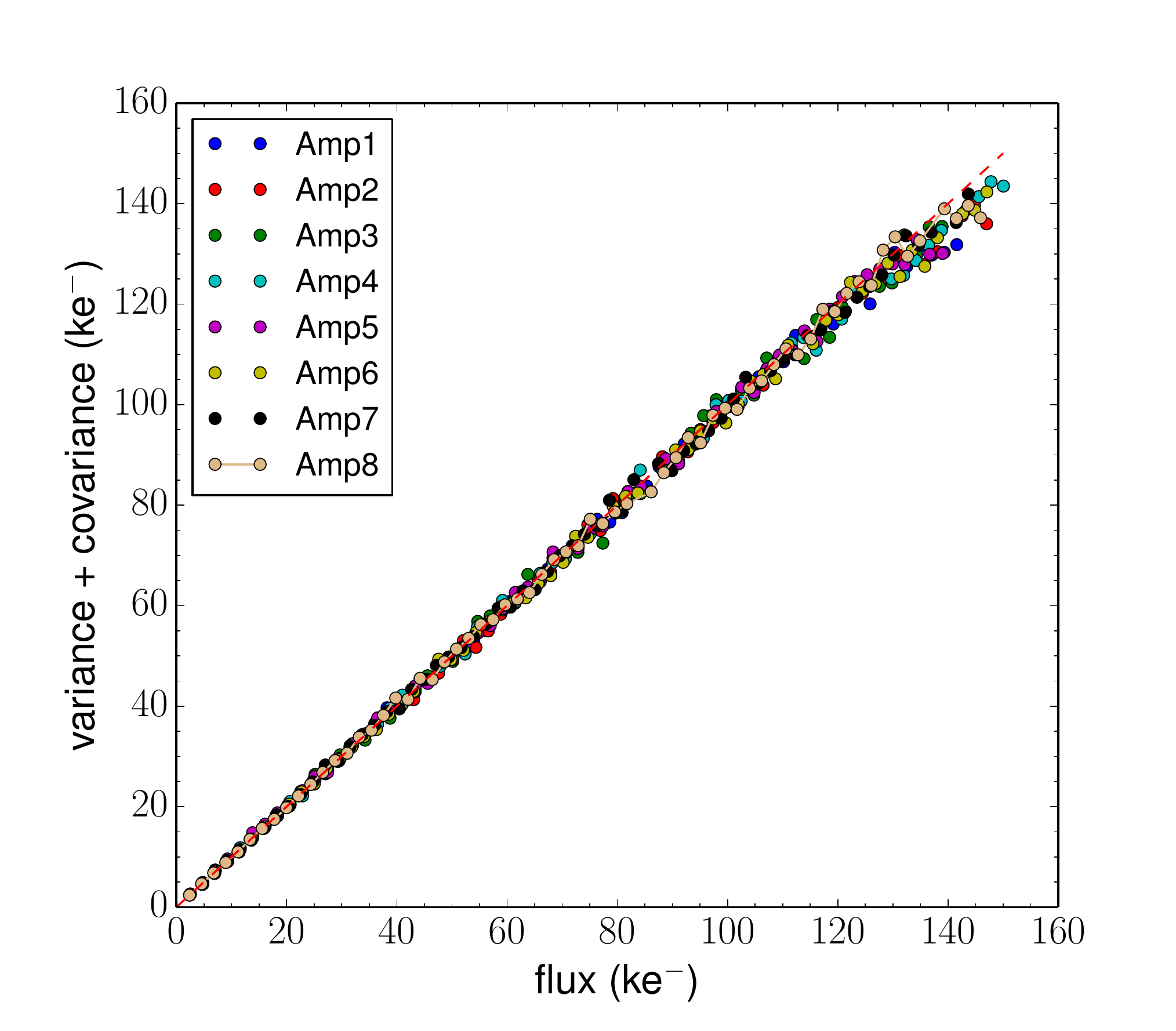}
\caption{Comparison for the CCD E2V-250 between the expected 
Poisson noise (red dashed line) and raw 
PTCs that are corrected by summing covariances up to 4 pixels distance.
For a 100 ke$^-$ flux level, these correlations add up to 18\% of the variance.
The corrected PTCs slopes coincide with Poisson law at $\approx$ 0.5\%, indicating
that more than 97\% of the correlations are considered by the
truncation at a 4 pixels distance of the integral of the correlation function. 
  \label{fig:varprof}}
\end{figure}

\subsection{Summary of correlation properties}

Linearly increasing correlations are seen on all tested CCDs.
They are the origin of the quadratic behavior of the PTC.
It requires care to precisely separate them from 
 other already identified correlating processes, but then
they are detected up to 4 pixels distance at a level of
a few 1$\times$ 10$^{-4}$. We propose in the next section a physical source 
of these correlations that is also found 
to generate a broadening of spot-like illumination.

\section{A simple model of Coulombian forces within CCDs}
\label{coulomb}

We propose an explanation for both the brighter-fatter effect and correlations in 
flatfields, which involves transverse 
field line displacements due to charge 
distribution within surrounding pixels.
With a simple electrostatic simulation, we evaluate 
how much it would affect spot broadening and pixel
correlations. We further
derive a simple model that uses correlation measurements to predict brighter-fatter
relations.

\subsection{Evolution of electrostatic fields as charges accumulate into pixels}
\label{elecsim}

Combining the observation of
correlations in flatfield images with the broadening of
stars with flux, we picture the effect that the charges accumulated in a CCD
perturb the drift electric field that subsequent charges will
experience. These perturbations tend to drive drifting charges away
from pixels {with higher counts than their surroundings. 
In flatfield images, these higher counts result from Poisson
fluctuations, while they result from genuine illumination variations in
star or spot images.

The relevance of this description has already been assessed in \citet{BNL13}
by showing that a simple electrostatic simulation of the pixels
reproduces both the brighter-fatter effect and the scale of pixel correlations.
Figure \ref{fig:electric_field} illustrates the phenomenon  at play by superposing  
the electric field lines for empty pixels and for a  pixel that is filled with 50 ke$^-$.
In this simulation, the pixel geometry correspond to the CCD E2V-250, and we 
approximate the intrinsic silicon
as free of charges. The CCD is simulated with a bias voltage (BSS) of 70 V, 
a clocking voltage (CV) of 10 V, and a depth where charges 
accumulate of \SI{2.5}{\micro\metre}. The same electrostatic 
potential that separates pixels in rows is applied to the column separation 
(because we do not know the exact profile
of the implants that define the column separation). 
The pixels boundaries are found by following field lines from the top
to the bottom of pixels. The figure illustrates that these boundaries are
displaced by the charge pattern stored in the device. The sense of the
effect (due to repulsion of same sign charges) is that pixels with higher counts 
than their surroundings shrink as they fill up, while their neighbors widen and
slightly shift. The figure also illustrates that the second neighbor
pixel is also affected.
The consequence of this evolution of the
pixel effective size\footnote{The accumulation of charges also modifies
the effective size of the pixel because of an attenuation of the longitudinal 
component of the electric field, which causes an increase of the diffusion.
This contribution is discussed in \S \ref{ids}.}
is that pixel correlations
in flatfield images should increase with increasing fluxes and that
point sources should appear broader as they get brighter. The photon noise that
contributes to the contrast of a flatfield is reduced as drifting charges are being more repelled
by a pixel where more charges have already accumulated.
Meanwhile the perturbation of field lines
in the surrounding of a spot in an astronomical image results in a broadening 
of its width as the source becomes brighter.

The agreement that was found between the simple electrostatic simulation
and the observations is only qualitative, and a 
quantitative prediction would
necessitate a detail knowledge of the geometry of the CCD and 
of the doping that are not
available. To test the quantitative prediction, we have developed a 
generic model that describes both effects 
with the same algebra.

\begin{figure}
\centering
\includegraphics[width=0.96\linewidth]{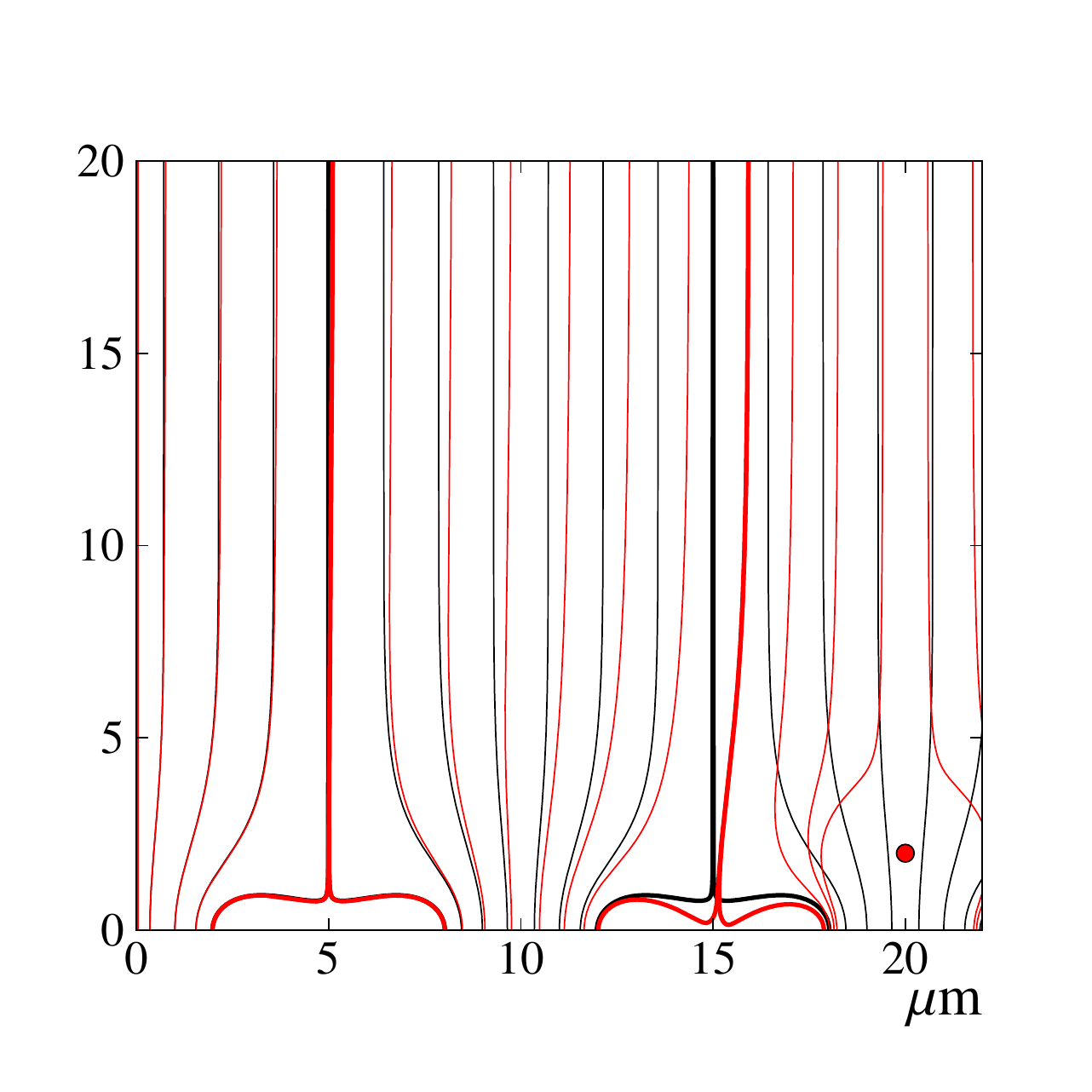}
\caption{Electrostatic calculation of Coulombian forces generated by the electrode 
voltage (CV), the depletion voltage (BSS), and the electrons collected. 
This figure is a zoom of the last \SI{20}{\micro\metre} of \SI{100}{\micro\metre} 
thick simulated pixels. The black lines represent the electric field for empty pixels, 
and the red lines represent the electric field when the right most pixel is filled with 50 ke$^-$. 
The separations between two adjacent pixels are indicated by the bold lines and are shifted to
 the right when adding the charges. This results in a variation of pixel 
effective size: the size of the filled pixel has decreased,
while the size of neighbor pixel has increased (and its centroid  has slightly shifted). 
This figure also qualitatively confirms the observation that the effect is 
achromatic: we see that drift trajectories are altered in the bottom 
$\sim$ \SI{10}{\micro\metre} of the device only, and hence do not depend 
significantly on the conversion depth.
 \label{fig:electric_field}}
\end{figure}

\subsection{Empirical parametrization of the pixel size variations as a function of flux}
\label{sec:algebra}

In this section, we derive a parametrized model
of the effects of electric field distortions in a CCD
that are induced by the charges residing within the CCD during the exposure.
We model the displacement of the effective boundaries of a pixel 
(labelled $(0,0)$), which is caused by a charge $q_{i,j}$ in a bucket at position $(i,j)$ as
\begin{equation}
\delta^X_{i,j}/p = a^X_{i,j} Q_{ij}/2 \; ,  \label{eq:def-a}
\end{equation}
where we have expressed that the (perturbing) electric field 
due to a charge is proportional to this charge ($Q_{ij}$), 
and approximated
alterations to drift trajectories to first order.
The variable $p$ refers to the pixel size, and we have introduced a factor of two for convenience.
The variable $X$ indexes the four boundaries of the pixel $(0,0)$,
and we label each boundary by the coordinates of the pixel
that shares it with (0,0): $X \in \{(0,1),(1,0),(0,-1),(-1,0)\}$.
The $a^X_{i,j}$ coefficients that define the model\footnote{Compared to \cite{BNL13},
we have chosen a different normalization of the $a^X_{ij}$ coefficients here, because 
the relation between electrostatic calculations and predicted effects is now straightforward.}
 satisfy symmetries
\begin{align}
a^X_{i,j} &= a^{-X}_{-i,-j}   & \text{(parity)} \label{eq:par}\\
a^{0,1}_{i,j} &= -a^{0,-1}_{i,j-1}\; . & \text{(translation invariance)}\label{eq:trans}
\end{align}

Each boundary of the pixel $(0,0)$ shifts under the influence
of all charges. Its displacement reads
\begin{align}
\frac{\delta^X}{p}  & = \sum_{i,j} \delta^X_{i,j}/p \\
          & = \frac{1}{2} \sum_{i,j} Q_{i,j}  a^X_{i,j}\; .
\end{align}
If all charges $q_{i,j}$ are equal, the electric field induced on a pixel 
boundary vanishes, and the boundary does not shift.
As a consequence, the  $a^X_{i,j}$ have to obey sum rules:
\begin{equation}
\sum_{i,j} a^X_{i,j} = 0 , \forall X \; . \label{eq:sum-rule}
\end{equation} 

We call charge transfer 
the difference between charge contents with and without the
perturbing electric fields. 
The displacement of the pixel boundary $\delta^X$ induces a charge transfer
between the pixel $(0,0)$ and the pixel $X$, where $X \in$ 
$\{(0,1),(1,0),(0,-1),(-1,0)\}$. To first order, this charge
transfer is proportional to both the pixel boundary displacement
and to the charge density flowing on this boundary. For a well-sampled image
(i.e. the charge distribution impinging a pixel does not vary rapidly
within this pixel), we can approximate the charge density drifting
on the boundary between pixel $(0,0)$ and its neighbor $X$ 
as 
\begin{equation}
\rho_{00}^X = (Q_{0,0}+Q_X)/2 \; ,\label{eq:density-approx}
\end{equation}
so that the net charge transfer due to perturbing electric fields
between pixel $(0,0)$ and its neighbor $X$  reads
\begin{align}
\delta Q^X_{0,0} & =  \frac{\delta_X}{p} (Q_{0,0}+Q_X)/2 \nonumber \\
                & =  \frac{1}{4} \sum_{i,j} a^X_{i,j} Q_{i,j} (Q_{0,0} + Q_X)\; . \label{eq:delta-q-x}
\end{align}
The expression is non-linear with respect to the charge distribution:
the charge $Q_{i,j}$ is the source charge, and the expression
$(Q_{0,0} + Q_X)$ approximates the test charges. 
A calculation of $\delta Q^X_{i,j}$ boundary displacements
resulting from a spot having 100 ke$^{-}$ in pixel maximum and a Gaussian shape with
a RMS of 1.6 pixel
is given as an illustration figure \ref{fig:es}.
The net charge transfer within the central pixel results in a decreasing 
effective size, while the effective size of pixels away from spot center tend to grow.  

\begin{figure}
\centering
\includegraphics[width=1.0\linewidth]{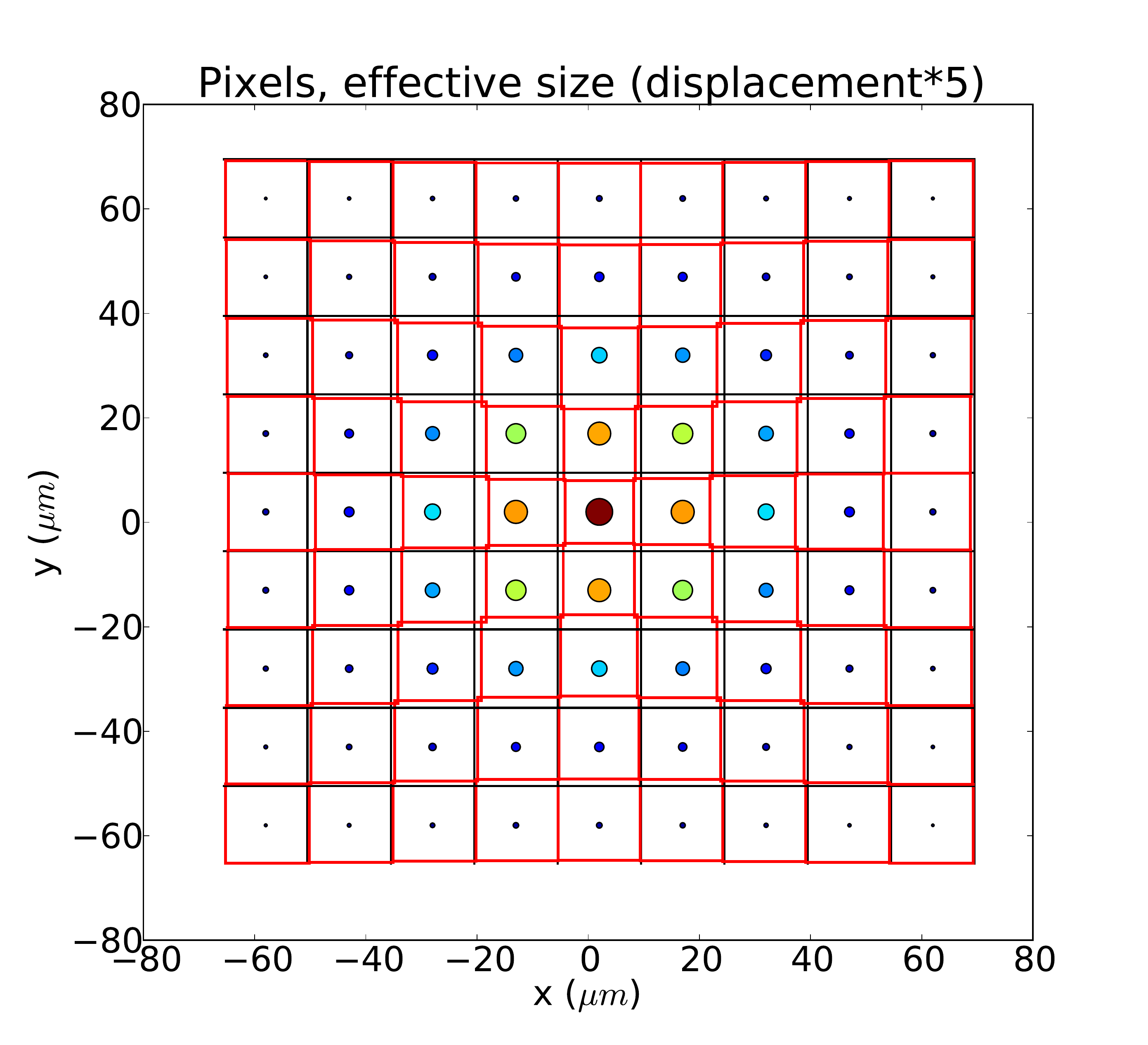}
\caption{Illumination with a spot having 100 ke$^{-}$ in pixel maximum
and a Gaussian shape with a 
r.m.s of 1.6 pixel results in boundary displacements (multiplied by a factor
5, red lines) that depend on the 
distribution of charges within surrounding pixels (colored circles).
The displacements between the intial boundaries (black)
and their final positions (red) correspond to the $\delta^{X}$ terms
of our model.
It shows that the central pixel effective size shrinks as illumination increases,
while pixels away from spot center tend to grow.  
  \label{fig:es}}
\end{figure}

When dealing with electrostatic simulations, we have to account
for the charge build-up during integration: on average the perturbations
to the drift electric field are
over the exposure time half of what they are at the end of the exposure.
This justifies the factor of 2 in expression \ref{eq:def-a},
so that the $a^X_{ij}$ coefficients can be extracted from simulations
as the ratio of the boundary displacement (in pixel size unit) to the source charge.

The perturbed charge in pixel $(0,0)$ reads
\begin{equation}
Q'_{0,0} = Q_{0,0} +\delta Q_{0,0} \label{eq:perturbPTC}
\end{equation}
with 
\begin{equation}
\delta Q_{0,0} = \sum_X \delta Q^X_{0,0} = \frac{1}{4}\sum_X \sum_{i,j} a^X_{i,j} Q_{i,j} (Q_{0,0} + Q_X)\; . \label{eq:scrambling_model}
\end{equation}

To evaluate the statistical correlations introduced
by the charge-induced perturbations of drift trajectories, we wish
to evaluate 
\begin{align}
Cov(Q'_{i,j}, Q'_{0,0}) &= Cov(Q_{i,j}, \delta Q_{0,0}) + [(i,j)\leftrightarrow(0,0)] + O(a^2) \nonumber \\
                     &= 2Cov(Q_{i,j}, \delta Q_{0,0}) +  O(a^2)\; , \label{eq:cov_2}
\end{align}
where $O(a^2)$ stands for expressions quadratic in the $a^X_{i,j}$
coefficients of expression \ref{eq:delta-q-x}. We stick to
first order perturbation expressions, as real data indicates that
this is justified. In the case where the pixel $(i,j)$ is not a nearest 
neighbor of $(0,0)$, we have
\begin{equation}
Cov(Q_{i,j}, \delta Q_{0,0}) = \frac{1}{4} Var(Q_{i,j}) \sum_X a^X_{i,j} E[Q_X+Q_{0,0}]\; ,
\end{equation} 
which for a flatfield illumination of average content
$\mu$ and variance $V$ reads,
\begin{equation}
Cov(Q_{i,j}, \delta Q_{0,0}) = \frac{1}{2} V \mu  \sum_X a^X_{i,j}\; .
\end{equation}
In case the pixel $(i,j)$ is a nearest neighbor of pixel $(0,0)$, say $Y$, we 
have two terms in the covariance\footnote{The exact expression involves the third moment
(or the skewness) of the charge probability density function, which is non-zero for a Poisson distribution. 
However, since we can approximate the statistics of pixel content with a Gaussian distribution,
which has no skewness, this contribution of the third moment can be safely neglected. The
relative error is of order $1/<Q>$ with $Q$ expressed in electrons.} :
\begin{multline}
Cov(Q_Y, \delta Q_{0,0}) \simeq  \\
 \frac{1}{4} Var(Q_Y) \sum_X a^X_Y E[Q_X+Q_{0,0}] + \frac{1}{4} \sum_{i,j} a^Y_{i,j} Var(Q_Y) E[Q_{i,j}]\; .
\end{multline}
For a flat, the second term vanishes because of the sum rule (eq. \ref{eq:sum-rule}). So, using Eq. \ref{eq:cov_2}, we find that whether or not $(i,j)$ is a nearest neighbor,
the covariance between pixels in a uniform exposure of average $\mu$ and variance V reads
\begin{equation}
Cov(Q'_{i,j}, Q'_{0,0}) = V \mu  \sum_X a^X_{i,j} \label{eq:covar}
\end{equation}
to first order of perturbations. The electrostatic influence from
collected charge, thus,  induces covariances between pixels in uniform
exposures that scale with the average and the variance of
pixel contents. If one measures correlation coefficients (ratio of
covariance to variance), those are expected to scale with the 
illumination level of the uniform exposure. In the same manner, applying this equation
to $(i,j)= (0,0)$ and replacing in the expression \ref{eq:perturbPTC}, it shows that our 
model of electrostatic influence also predicts a quadratic behavior of the PTC :
\begin{equation}
Cov(Q'_{0,0}, Q'_{0,0}) = V +  V \mu  \sum_X a^X_{0,0} \label{eq:quadptc}\; ,
\end{equation}
 which allows us to identify the $\alpha$ term from the equation \S \ref{ptc} as a combination
of the $a_{00}$ parameters:
 \begin{equation}
   \alpha = - \sum_X a^X_{00}\; . \notag
 \end{equation}
We notice that $\alpha$ is positive because the four $a^X_{00}$ terms are negative 
(they correspond to the narrowing of a pixel due to one collected charge within).

\citet{ma}
propose a model, which corresponds to a peculiar solution of eq. \ref{eq:covar} 
where all the $a^X_{i,j}$ of a 
given pixel are set equal. This parametrization 
totally contradicts the findings of \S \ref{elecsim}
(see for instance figure \ref{fig:electric_field}). They refer to their model as 
a``charge-sharing'' phenomenon. It follows a proposition 
from \citet{Downing13}, who report a variation of the non-linearity of the PTC when varying the 
collection phase voltage and who interpret it as a consequence of the variation 
of the lateral diffusion. It should be noted that lateral diffusion variation
cannot be evocated without the fact that mean trajectories are also modified. 
We show in the next section that this diffusion is a sub-dominant contribution for the 
brighter-fatter effect.

\subsection{Influence of diffusion in the substrate}
\label{ids}

Although diffusion actually smooths charge distributions, it is not
the best candidate to explain the brighter-fatter effect or
correlations in flatfields. The diffusion equation is linear and
hence is not obviously well suited for describing a linearity-violating
phenomenon, such as the brighter-fatter effect. Diffusion also does not
cause correlations between pixels as long as the probability to
convert in a given pixel and be collected in some other pixel does not
depend on the pixel contents. These are first order arguments because
diffusion indeed contributes to both the brighter-fatter effect and
correlations and because the charge contents alter (indeed reduce) the
longitudinal drift electric field on which the diffusion coefficient
depends. The mechanism is thus the following (\citealt{holland13}): a
pixel with higher counts than its surroundings 
will have a lower drift electric field than its
neighbors (see e.g. \citealt{Kent73}), and the increased diffusion
will increase in turn the probability of transferring photo-electrons
to neighboring pixels.

The impact of diffusion to transfer charges across pixels can be
sketched to first order of perturbations: a charge stored in the CCD
will alter the probability that electrons converted above a pixel
diffuse to a neighboring pixel. To first order, this probability is
proportional to the source charge, and the amount of transferred charge
via this mechanism follows the expression \ref{eq:delta-q-x}, although
we had invoked boundary displacements to justify it. So, our model
from \S \ref{sec:algebra} is actually fully compatible with this
second order effect of diffusion. We, hence, expect that the latter does
not break the relation established by our model between correlations
and the brighter-fatter slope. We can, however, readily note that since
the contribution of diffusion depends on the conversion depth (which
varies on average with wavelength), the fact that we do not observe a
significant chromatism of the effects indicates that this second order
contribution of diffusion is modest. It is nevertheless interesting
to evaluate its contribution to the overall broadening of spots or
stars, for example to assess the impact of ignoring it in
the outcome of electrostatic simulations.

We have hence evaluated the contribution
of the variations of diffusion induced by the charges stored in the
CCD  to the brighter-fatter effect, using the following assumptions:

\begin{itemize}
\item[$\bullet$]  Lateral diffusion causes a spot broadening that is $\propto \sqrt{2Dt_{r}} $,
where $t_{r}$ is the transit time.
\item[$\bullet$]  Transit time is estimated by $t_{r}=(1/\mu)\int_{d+}^{Z}{dy/E(y)}$, 
assuming that charges velocity is $v = \mu E(y)$.
\item[$\bullet$]  The variation of the ``instrumental'' PSF ($\sigma_{PSF} = \sqrt{2Dt_{r}} $) 
is then computed 
using our electrostatic 
 simulation starting with empty pixels and adding charges up to 100 $ke^{-}$.
\item[$\bullet$]  For DECam, we use its nominal backside substrate voltage and clocking voltage parameters: 
a BSS of 40 V and a CV of 6.5 V. 
For CCD E2V-250, we use the values applied to acquire these 
data sets: a BSS voltage of 70 V and a clocking voltage of 10 V.  
The pixel thickness (Z) and width are as indicated by manufacturers. The depth
of the buried channel, which is the only free parameter ot the simulation, is set to \SI{2}{\micro\metre}.
\end{itemize}

Our results are summarized in table \ref{tab:diffcont}. For a device operated at low drift field, such as DECam (E=0.16
V/\SI{}{\micro\metre}), we find that the variation of lateral diffusion is $\sim$20\%.
Given that the contribution to the broadening is half the maximum variation, it contributes 
to $\sim$10\% of the
brighter-fatter slope (for realistic IQ conditions), while the contribution is no more than
a few \% of the observed brighter-fatter slope for the
higher drift field (E=1 V/\SI{}{\micro\metre}) that is used to operate the 
CCD E2V-250. So, this effect alone
does not quantitatively match the measurements reported in \S \ref{Results}. It even
appears to be secondary when simulating ``high-field'' devices 
(such as presented in \S \ref{elecsim})
for which simulations accounting for the alterations to the average drift trajectory 
might be sufficiently accurate. In any case, the applicability of the parametrization
that we present in 
\S \ref{sec:algebra} does not depend on a detailed description which separates 
the relative contributions of lateral and of longitudinal electric fields 
to the charge transfers: the outcome of our correction method is relevant for a physical
interpretation that is a combination of both diffusion and drift of the electrons.

\begin{table}
\caption{Upper limit of diffusion variation for CCD E2V-250 
illuminated by 550 nm spots and for DECam CCD N17 in $r$-band.\label{tab:diffcont}}
  
  \begin{tabular}{c|cc|cc}
    \hline
    \hline
               size (\SI{}{\micro\metre})      &\multicolumn{2}{c|}{CCD E2V-250}&\multicolumn{2}{c}{DECam} \\
                               &    X        &  Y                    & X               &  Y   \\
\hline
\hline
Initial PSF             &  15.94      &  16.22           & 25.64   &  28.86 \\
PSF at 100 ke$^-$        &  16.41      &  16.74           & 25.97   &  29.18\\
Observed increase        & 0.47 &  0.52           & 0.33   &  0.32\\
\hline
Diffusion ($\sigma_{PSF}$) &  <4.00      &  <4.00           & <7.00   &  <7.00  \\
\begin{minipage}[c]{0.35\linewidth}\begin{center}Diffusion induced\\increase at 100 ke$^-$\end{center}\end{minipage} 
                         &  0.018       &  0.018          & 0.067   &  0.067 \\
\hline
\hline
Diffusion contribution (\%) &  3.7         &  3.4            & 20.2   &  20.7   \\
\hline
\end{tabular}
\tablefoot{The relative difference between the two results 
is coming from the operating bias voltages (CCD E2V-250  is operated with a $\sim$6 
times higher drift field than DECam). 
The final contribution of the diffusion to the PSF broadening is half its total 
variation.}
\end{table}

\section{Connecting flatfield correlations with the broadening of spot-like illumination}
\label{sec:comp}

For CCD E2V-250, DECam, and MegaCam, we compare the
  brighter-fatter slopes expected from the correlations measured in
  flatfield exposures with the brighter-fatter slopes measured
  directly. The first step is to derive the values of the
 model coefficients $a^X_{i,j}$ from the correlations. By using the found coefficients, the second step is
 to emulate the electrostatic distortions by transforming images of
 faint spots into realistic bright spots and compare those with real bright spots. 

\subsection{Extracting the model coefficients}
\label{sec:extracting_coefficients}

The $a^X_{i,j}$ coefficients are determined from the correlation coefficients $R_{i,j}$. 
They are extracted from correlation slopes
fitted on flatfield pairs up to PTCs extremum.
We have measured correlations up to 4 pixel separation,
which is 25-1 measurements, or $n^2-1$ for $n=5$.
Each pixel has 4 boundaries; there are $4 \times n^{2}$
coefficients $a^X_{i,j}$ to be determined. With $n=5$, there are 100 boundaries
to  consider. The internal consistency of the model directly removes 
the calculation of 40 boundaries that are shared by two pixels. The parity of the effect 
(Eq. \ref{eq:par} in \S \ref{par:pix}) also removes ten boundaries that are mirror 
of one another as seen from the source charge (5 in $i$ and 5 in $j$).
So we are left with $2 \times n^{2}$ (50) coefficients to evaluate.

There are $n^2-1$ measured correlations (related to
coefficients by Eq. \ref{eq:covar} in \S \ref{par:pix}),
and we complete those by $n^2+1$ extra constraints in order
to have a closed system. We derive the needed constraints from
smoothness considerations: the measured correlations decay
smoothly with distances and so should the $a^X_{i,j}$ coefficients.
We will first derive a crude model for these coefficients from the 
measured correlations, and then use this model to constrain ratios of
$a^X_{i,j}$ at similar distances, and constrain  
values at the farthest boundaries. We then directly solve and assess
how our predictions for the brighter-fatter slopes depend on the intermediate
smoothing model used. 

The smoothing model assumes that $a^X_{i,j}$ coefficients are the
product of a function of distance from the source charge to the considered
boundary ($r_{ij}$) and that it also trivially depends on the angle between the
source-boundary vector and the normal to the boundary ($\theta^X_{i,j}$):
$$
a^X_{i,j} = f(r_{ij}) \cos{\theta^X_{i,j}} \; .
$$
We have tried various analytic forms for $f(r)$ and settled for the exponential integral function:
\begin{align}
f(r) & = p_0 Ei(p_1 r) \label{eq:f_r} \\
Ei(x) & \equiv - \int_{-x}^\infty \frac{e^{-t}}{t} dt \; , \nonumber
\end{align}
where $Ei$ is only defined for $x>0$ and the singularity
of the integrand around $t=0$ is handled by taking the principal part.
We determine $p_0$ and $p_1$ by least-squares:
\begin{equation} 
\chi^2 = \sum_{i,j}\left(\frac{Cov_{ij}}{ V \mu} - \sum_X \left(p_0 Ei(p_1\cdot x^X) \cdot cos\theta^X_{ij}\right)\right)^2 \; .\nonumber
\end{equation}
Because we observe anisotropic correlations at short distances (e.g.
$Cov_{01} \neq Cov_{10}$), we do not include the three nearest neighbors
in the fit. The fit to the 21 other correlations is shown on Fig.~\ref{fig:ccl}
 with the minimization values. The agreement justifies the choice
of the exponential integral function.

\begin{figure}
\centering
\includegraphics[width=0.9\linewidth]{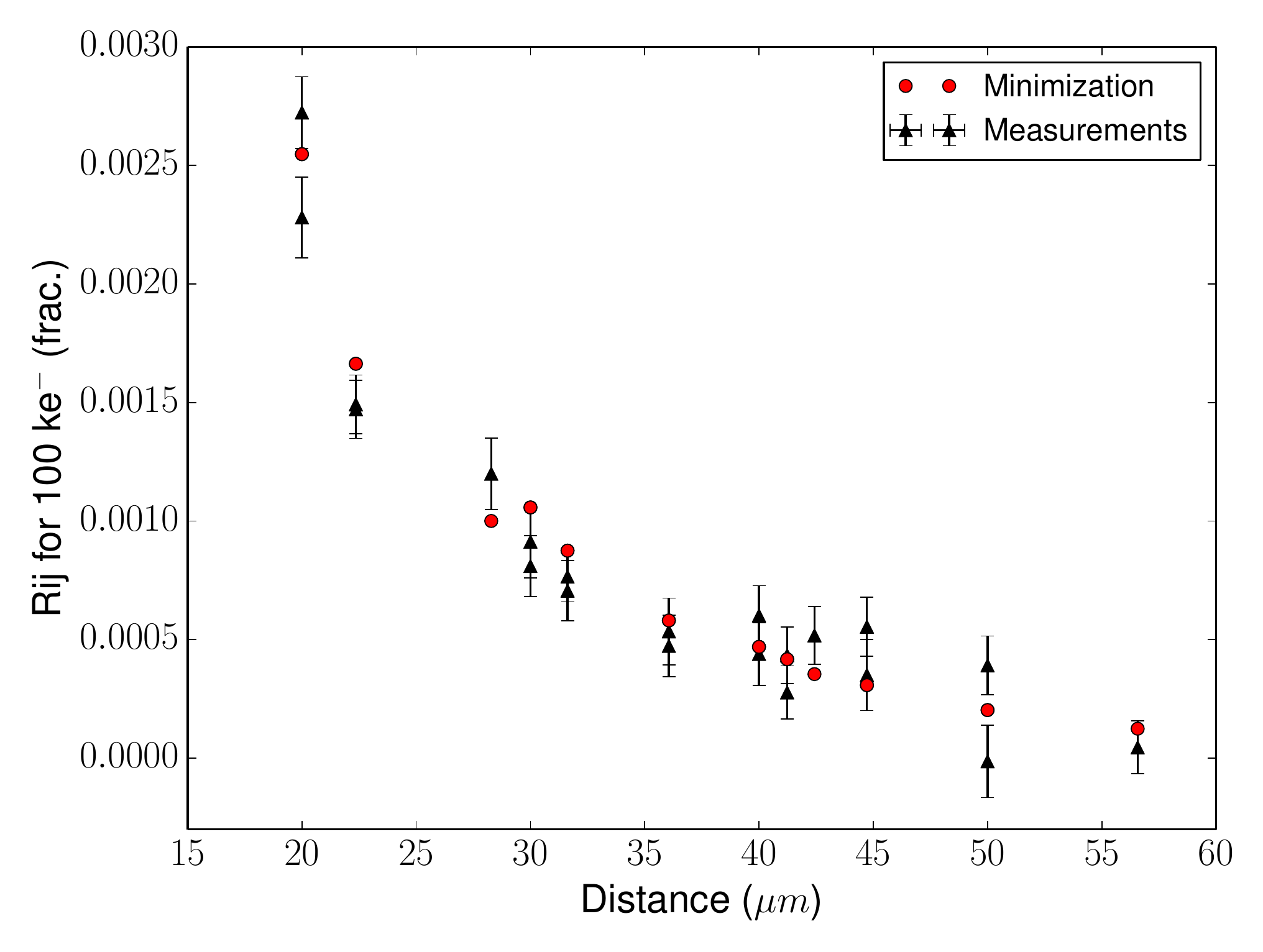}
\caption{Correlation coefficients of the non-nearest neighbors for the CCD E2V-250 
 are represented as a function of their distance (black triangles) and the correlation coefficients reconstructed from a $\chi^2$ minimization (red circles) using an Exponential Integral function. This function is used to estimate
 the last boundaries displacement.
\label{fig:ccl}}

\end{figure}

We use the fitted $p_0$ and $p_1$ parameters to impose the values on
the farthest boundaries ($2 \times n$ constraints) and also to impose
ratios of coefficients addressing adjacent boundaries 
of the off-axis coefficients (i.e. $i,j>0$). It reads
\begin{equation}
a_{i,j}^{(0,-1)} = a_{i,j}^{(-1,0)} \left( \frac{ Ei(p_1\cdot r_{i,j}^{(-1,0)}) \cdot cos(\theta_{i,j}^{(0,-1)})}
{ Ei(P_1\cdot r_{i,j}^{(0,-1)}) \cdot cos(\theta_{i,j}^{(-1,0)})}\right)  \; . \notag  \\
\end{equation}
No constraints are 
applied on the displacement of the boundaries of the pixels (1,0) and (0,1) 
because the analysis of correlation properties 
presented in \S \ref{par:pix} indicates a different behavior for 
those than for the rest of the coefficients. The displacements of the boundaries 
that are next to the source charge 
are due to a combination of effects: some of them being unknown (the actual geometry 
of the collecting area), other being complex (a contribution both from charges diffusion 
and from electrostatic influence (\S \ref{ids})).
The solution of the model for all boundary displacements is represented by the red shaded area on figure
\ref{fig:df}. The width of the area corresponds to the propagation of the
 $\pm $1$\sigma$ uncertainty of the correlation coefficents measurement. It
exhibits an irregular variation with distance that is introduced by
the X/Y anisotropy in correlation coefficients, which is ignored
in our simple simulation. Nonetheless, the overall shapes are compatible. 
To assess the sentivity of the results to the smoothing model we vary $p_1$ by $\pm $5$\sigma$, which changes the $a^X_{i,j}$ coefficients by about $\pm 1\%$.
This in turn changes the brighter-fatter slope by about  
$3 \cdot 10 ^{-4}$.

\begin{figure}
\centering
\includegraphics[width=1.0\linewidth]{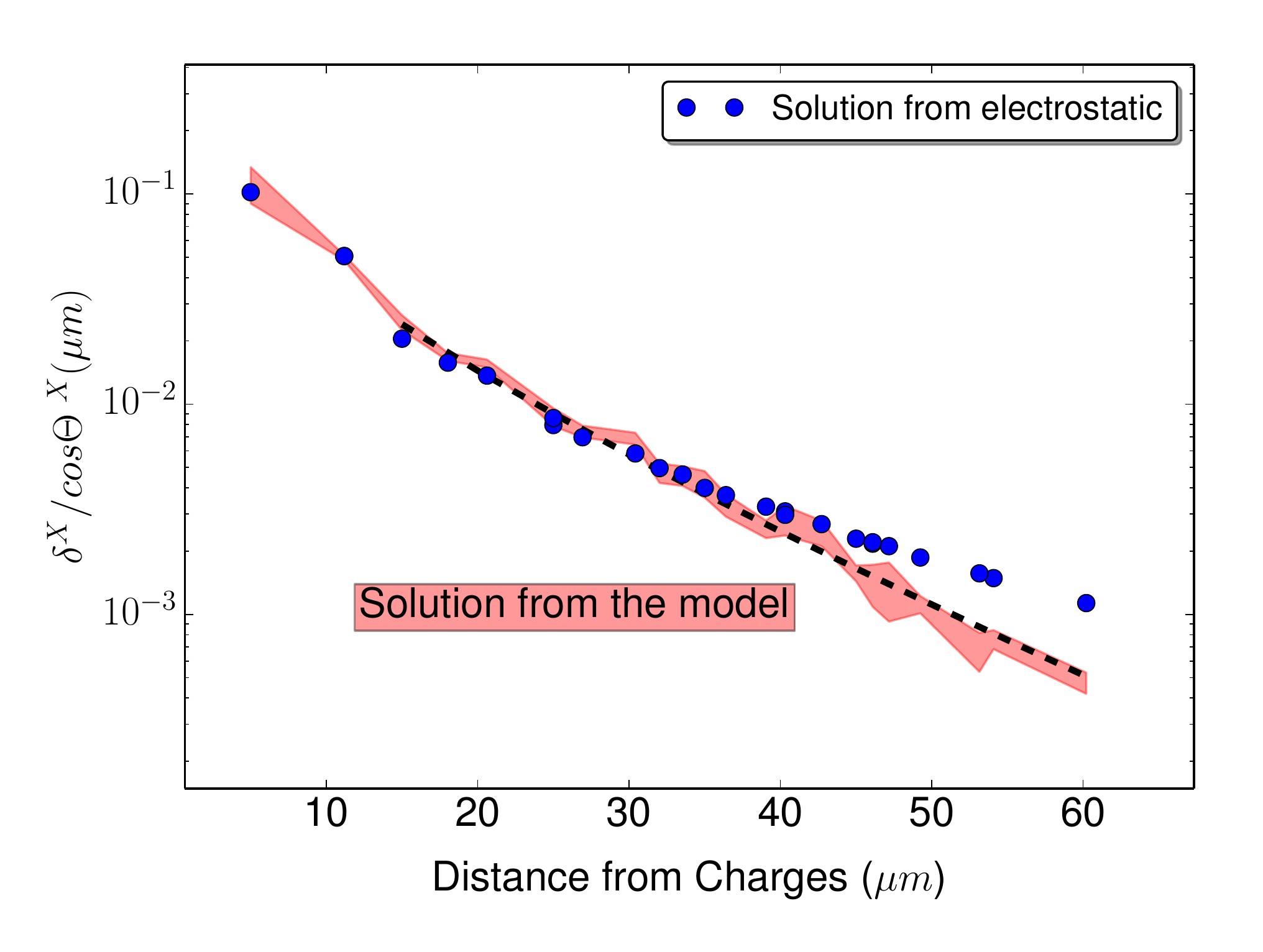}
\caption{Boundary displacements ($a^X_{i,j}\times $100 ke$^- $) for the CCD E2V-250 
projected in the direction radial from
the charges and represented as a function of the distance. We divide 
the coefficients by $cos\theta^X$ to display the 'radial
part' of those coefficients (the $f(r)$ function of Eq. \ref{eq:f_r}).
The result from the electrostatic 
simulation is indicated by the blue points, while the solution of the model with the propagation of the $\pm 1 \sigma$ uncertainties 
from the correlations measurement is indicated by the red shaded area. 
Its irregular shape is coming from the X/Y anisotropy in correlation 
coefficients, which is ignored in the simulation.  The dashed line indicates the
Ei function on the range where the constraint on pairs of adjacent pixels is used.
A  $\pm $5$\sigma$ variation of its $P_1$ parameter has an insignificant effect on 
the prediction of the brighter-fatter slope.
  \label{fig:df}}
\end{figure}

\subsection{Image sampling and estimates of perturbed charges distribution}
\label{Image sampling}

The last step to emulate the electrostatic distortion is to multiply 
boundary displacement $a^X_{i,j}$ with charge density 
flowing on boundaries. Assuming that the image is properly sampled, 
the density is approximated by interpolating among the pixel contents 
(equation \ref{eq:density-approx}). The accuracy of this approximation 
can be estimated by the simulation of spots with increasing 
size using Moffat functions. The exact charge density is then known everywhere in the images and, in particular, on the pixel boundaries.
The result of the test is shown in figure \ref{fig:bva} for the IQ ranging  
from 1.2 pixel to 3.5 pixels.
For an IQ of 1.6 pixel (such as our 550 nm spots on the CCD E2V-250), 
 with a relative brighter-fatter effect of $\sim$ 2$\%$, 
the approximation is introduces 
an underestimation on the size of the spot that is already below 0.1$\%$, which is less than 
5$\%$ of the amplitude of the effect.

\begin{figure}
\includegraphics[width=1.0\linewidth]{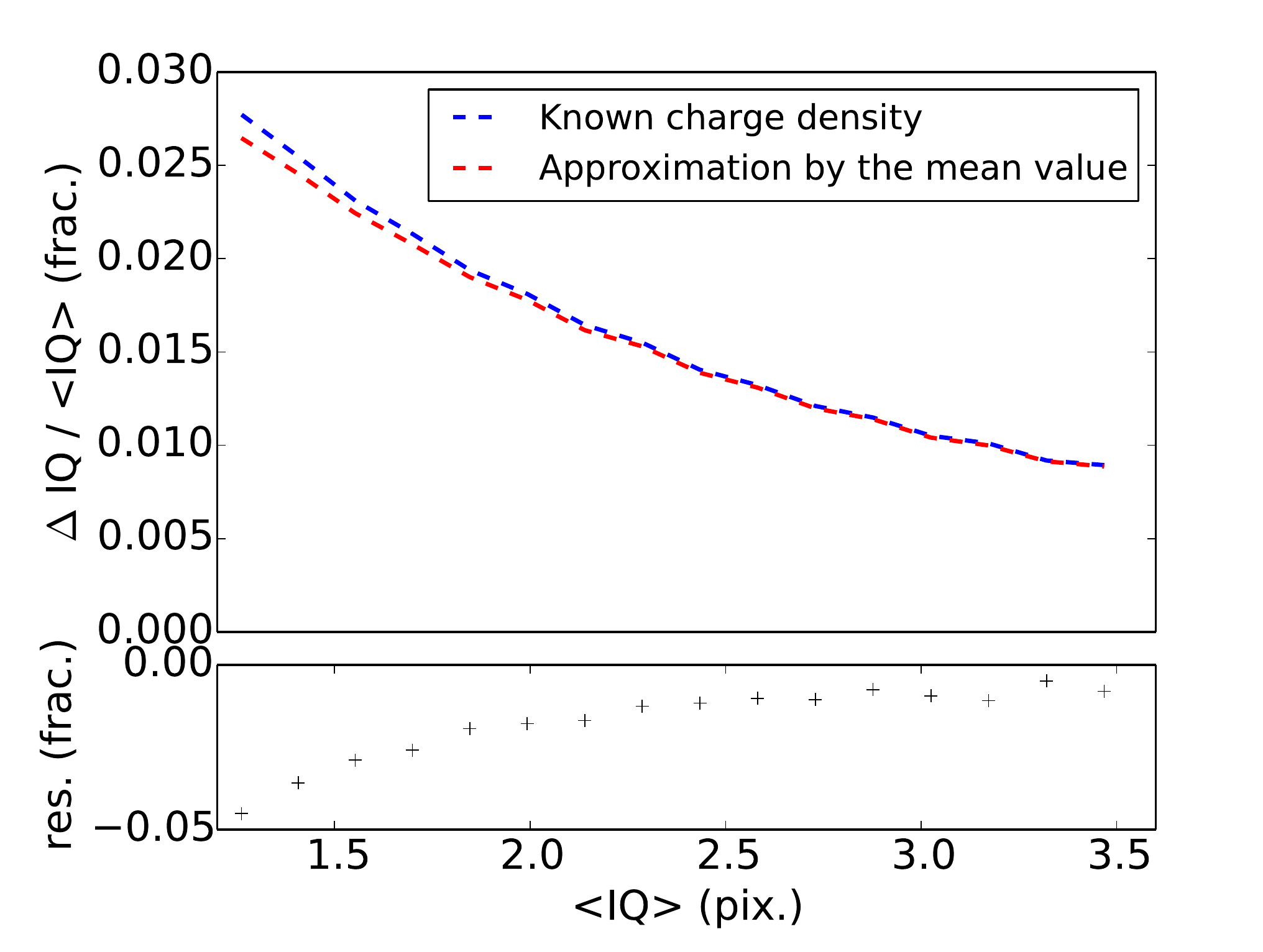}
\caption{Top panel: IQ variation for 100 ke$^{-}$ (peak) spots after having redistributed the
charges (on a range comparable 
to what was presented 
for DECam on figure \ref{fig:iqv}). The accuracy of the approximation 
of the interpolation method (in red) is compared with 
the result found using the exact charge density (in blue). 
Bottom panel: Residuals. For an IQ of 1.6 pixel, such as 550 nm spots on CCD E2V-250,
 the approximation introduced an underestimate below 0.1$\%$ on the size of the spot, which is less than 
5$\%$ of the amplitude of the effect.
  \label{fig:bva}}
\end{figure}

\subsection{Comparison of the redistribution model with the measurements}

The redistribution of charges in spot/star images consists of
\begin{enumerate}
\item Establishing a low flux spot/star model from images of faint spots 
in the case of CCD E2V-250 and a fitted PSF model using astronomical images
in the case of DECam and MegaCam.
\item Scaling up the image flux by the desired factor.
\item Transforming the image through 
  Eq. \ref{eq:scrambling_model} 
  (``redistribution'') from \S \ref{sec:extracting_coefficients}.
  
\end{enumerate}

\subsubsection{Redistribution of charges for the  CCD E2V-250 images}

\begin{figure}
\centering
\includegraphics[width=1.0\linewidth]{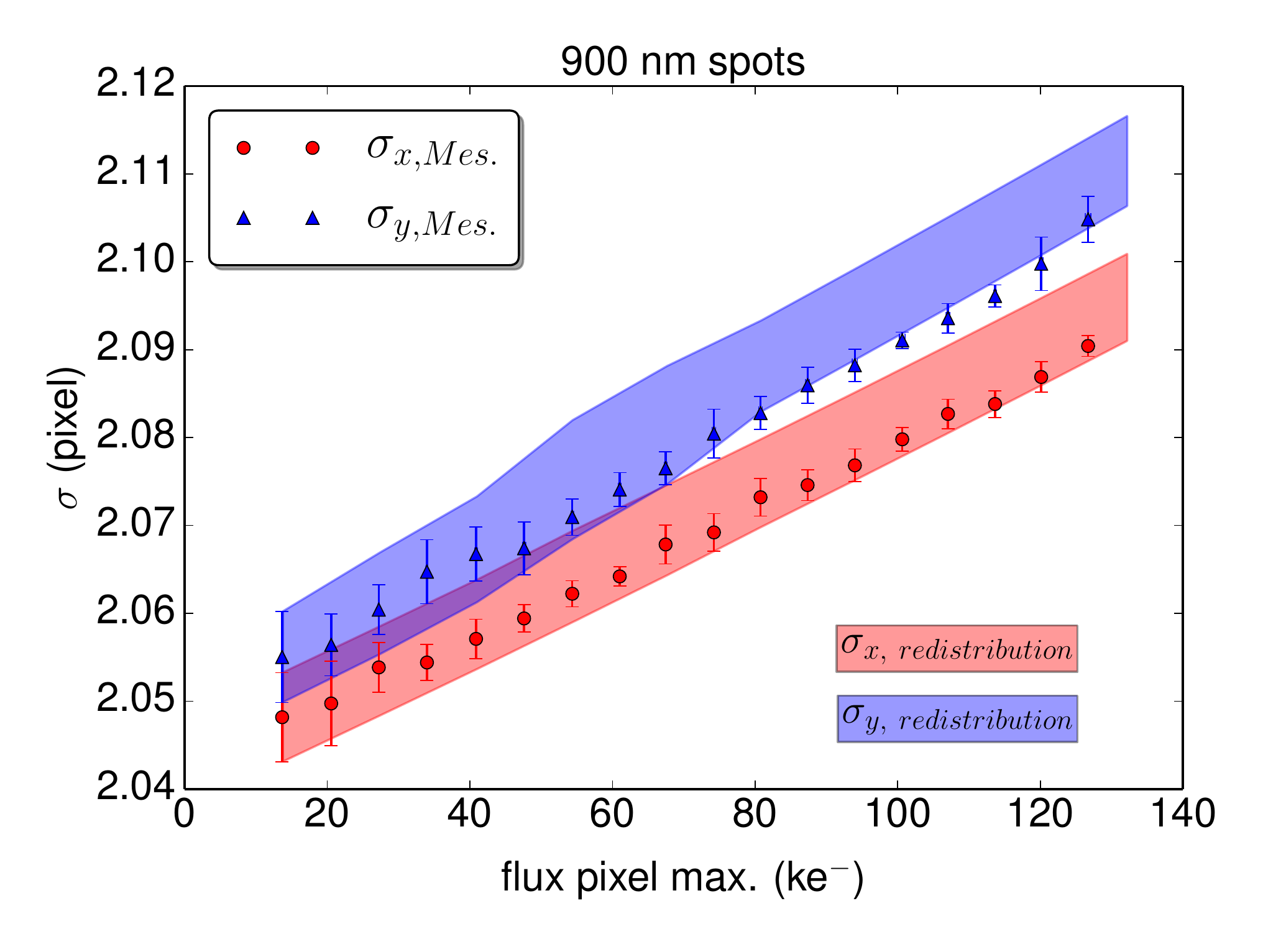}
\caption{Comparison between the broadening of spots from the redistribution
and the data (CCD E2V-250, 900 nm spots).   
The ten low flux spots ($\approx$ 13 ke$^-$) are redistributed after
increasing normalization to cover 
dynamical range (shaded areas). This shows the ability 
of our model of pixel effective size to 
reproduce the spot size increase observed in the brighter-fatter effect. 
  \label{fig:bf_lsst_e2v_900nm}}
\end{figure}

Redistribution of charges of low-flux 900 nm spots on the CCD E2V-250 does reproduce spot 
broadening with respect to increasing fluxes. This is illustrated in figure 
\ref{fig:bf_lsst_e2v_900nm}, where the redistribution of charges of low flux spots (shaded areas)
 is compared to the data (error bars represents $\pm$ 1$\sigma$ dispersion of measured second moments). 
It is seen that the trend is well reproduced both in X and Y directions. 
The constant width of the shaded areas originates in the initial dispersion 
of the 10 low flux spots (20 s exposure time, 13 ke$^-$ in maximum pixel).
As for the data, the dispersion of the measurement of the second moments
decreases as the signal-to-noise ratio improves at higher fluxes.

We can also approximate the inverse of the transformation of Eq. 
\ref{eq:scrambling_model} by using the same expression after flipping signs of 
the $a^X_{i,j}$ coefficients.
This transformation reduces spot widths down to their low-flux sizes. 
The accuracy of the transformation is assessed by evaluating the residual dependence of the  
spot width on the spot flux over the dynamic range. This is 
shown in figure \ref{fig:e2} for both 550 nm 
 and 900 nm spots. The uncertainty of the result is indicated and
separated between the statistical uncertainties that comes from correlation measurements 
and that are propagated using Monte Carlo simulations (plain 
colors) and the dispersion coming from second moment
 measurements (indicated in light colors). The dashed lines on the middle 
panel illustrates the results of reverse redistribution of charges when taking
correlations up to 1, 2, 3, 4 distance into account. It reduces the brighter-fatter effect 
by 20\%, 45\%, 70\%, and 87\%, respectively. The relatively small contribution of the correlations
from adjacent pixels emphasizes the importance of a long distance
mechanism contributing to the brighter fatter effect.
The bottom panel represents, for 
the 900 nm data set, the effect of the correction as a function 
of the distance in pixels, with the limit conditions for the $a^X_{i,j}$ taken into account. It indicates that
there is no further evolution farther than the 4 pixel solution. 

The slopes of spot broadening before and after reversing the broadening effect are 
summarized in table \ref{tab:dist_comp}. 

The broadening of the 550 nm spots is measured with a $\sim$3.6~\% relative precision on 
the X and Y slopes and  with a $\approx$1.5 \% relative precision  on the 900 nm spots.
The reverse redistribution method is found to consistently remove the spots broadening and leaves
no residual slopes to within 5\% of the initial effect. 
It is below 1$\sigma$ RMS of the combined uncertainties, with the exception of the Y direction 
of the 900 nm spots for which it is below 2$\sigma$ RMS
Although not significant, it should be pointed out that the amplitude of the residuals
are compatible with the expected underestimation introduced by the charge density approximation 
(see \S\ref{Image sampling}). Lastly, we observe from table \ref{tab:dist_comp} 
that for the 900 nm spots the errors on the correction are dominated by the model. It indicates
that the ability to measure distant correlations 
(down to 1$\times$10$^{-4}$ level) is an essential step for an accurate 
application of this brighter-fatter correction method.

\begin{figure}
\centering
\includegraphics[width=1.0\linewidth]{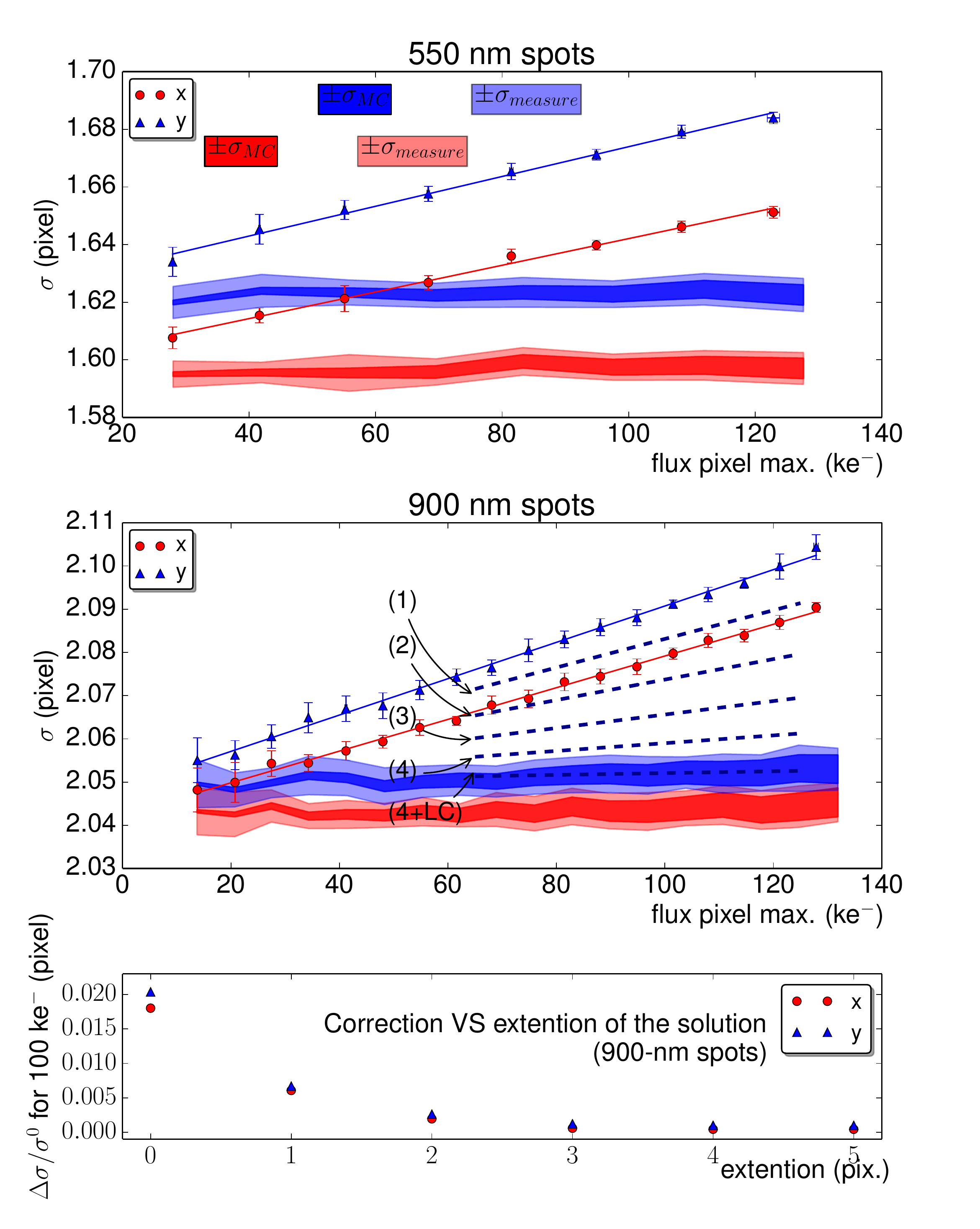}
\caption{Spot size along $X$ and $Y$ directions for 550 nm spots (top panel) and 900
  nm spots (middle panel) on the CCD E2V-250. Raw spots are fitted with lines
  and measurements after redistribution of charges
  with the inverse model are shown with shaded area. Light
  colors correspond to dispersion coming from spot size measurements 
  while darker colors represent the propagation of statistical 
   1$\sigma$ uncertainties on  correlations coefficients given that 50 
pairs of flats were used. 
  The five dashed lines on the lower panel indicate the redistribution prediction for
    the spot broadening in the Y direction when taking an increasingly large
    area of pixel correlations into account. It ranges from $\pm$1 pixels distances 
    to $\pm$4 pixels and $\pm$4 pixels plus the limit condition (4+LC) established 
in the previous section.
  In the latter case, it is found that the correction 
  restores the invariance of the PSF size with respect to increasing flux
  with a relative precision below 5$\permil$ at 550 nm and 3.4$\permil$ at 900 nm.
  The bottom panel represents the brighter-fatter effect as a function of
 	the extension of the solution with the limit condition taken into account. It shows
	that the boundaries displacement at distance further than 4 pixels have a negligible impact
	on the solution.
  \label{fig:e2}}
\end{figure}

\begin{table*}
\begin{center}
\caption{Parameters of linear fits of the data presented figure \ref{fig:e2}.
 \label{tab:dist_comp}}
\begin{tabular}{c|cc|c}
\hline
\hline
\multicolumn{1}{c|}{} &\multicolumn{2}{c|}{measurements} & \multicolumn{1}{c}{corrected} \\
\multicolumn{1}{c|}{Fit parameters}  & Slopes $\pm \sigma_{meas.}$  &  origin  & Slopes $\pm (\sigma_{meas.}\oplus\sigma_{model.}$)   \\
                &    [10$^{-4}$pix/ke]   &    [pix] &  [10$^{-4}$pix/ke] \\
\hline
\hline
X - 550nm &  4.61 $\pm$ 0.17 &  1.594  & 0.21 $\pm$ 0.23 \\
Y - 550nm &  5.06 $\pm$ 0.18 &  1.622  & 0.04 $\pm$ 0.20 \\

\hline
X - 900nm &  3.80 $\pm$ 0.05 &  2.042 & 0.15 $\pm$ 0.17 \\
Y - 900nm &  4.25 $\pm$ 0.06 &  2.048 & 0.22 $\pm$ 0.13 \\
\hline
\end{tabular}
\tablefoot{For the 550 nm spots, the statistic gives a $\approx$3.6 \% relative precision on 
the X and Y brighter-fatter slopes (first column). For the 900 nm spots, the higher statistic reaches a better relative precision ($\approx$1.5 \%) on 
the X and Y brighter-fatter slopes.
After the correction (second column), the residual slopes are below 5\% of their initial values.
 It should be pointed out that the amplitude of the residuals
are compatible with the expected underestimation introduced by the charge density approximation.
It is also found that these residuals are within the 1 $\sigma$ combined uncertainties from the measurements 
and the uncertainties from the redistribution of charges (except for the Y direction 
of the 900 nm spots for which it is below 2$\sigma$ RMS.)}
\end{center}
\end{table*}

\subsubsection{Redistribution of charges for DECam and MegaCam stars}

The redistribution method also reproduces the amplitude of the effect 
measured on DECam but with less precision due to the limited number
of flatfields in the  publicly available science verification images. 
A comparison is shown in figure~\ref{fig:n17} with CCD-S11,
 using a set of 20 $r$-band exposures acquired in December 2012
(giving $\approx$ 13 000 stars). A PSF model is extracted
from the set of images to serve as a spot reference. Charges are then redistributed 
after scaling up the spot reference in flux.
A Monte Carlo simulations is used to propagate 
uncertainties from the measurement of the correlations to
the prediction of brighter-fatter slope. On this CCD, it predicts a broadening of 
about 0.025 pixel between 0 and 100 ke$^{-}$, which is symmetrical
in X and Y direction. The RMS from coefficient measurement uncertainties 
is 0.004 pixel at  100 ke$^{-}$, 
which result in a 16\% relative uncertainty on the modeling 
of the effect. This is near twice as much as for CCD E2V-250,
and it is directly connected to our science verification 
data set that contains about four times fewer pairs of flatfields to estimate
correlation coefficients. Nonetheless, the average value of the redistribution method
reproduces the observations well, in both X and Y direction.

\begin{figure}
\centering
\includegraphics[width=0.9\linewidth]{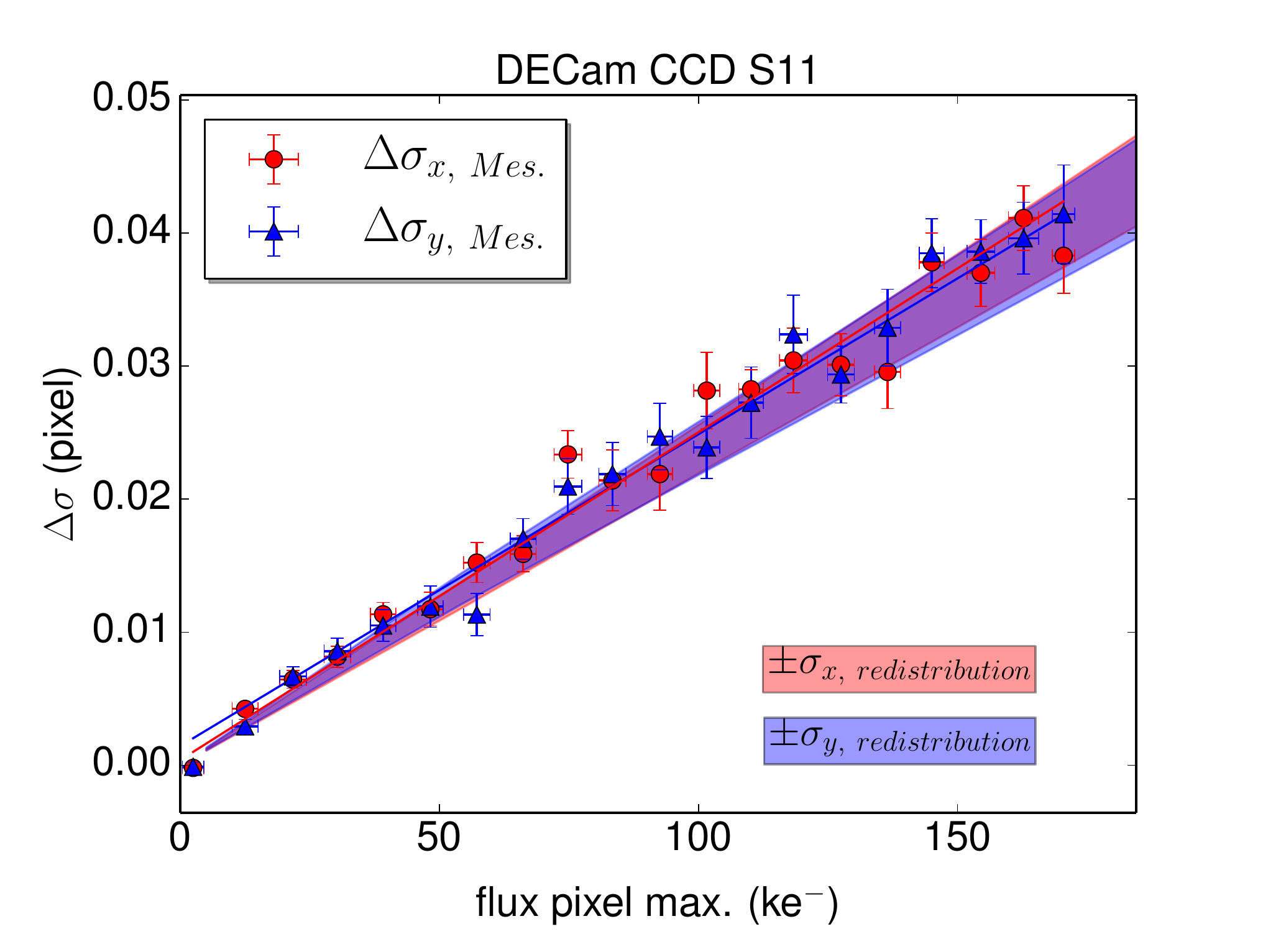}
\caption{Measured brighter-fatter effect and simulated effect seen on DECam CCD S11 and showing 
the propagation of correlation coefficient uncertainties. A $\pm$1$\sigma$ RMS 
is  0.005 pixel at 100 ke$^{-}$, giving a 25 \% relative uncertainty,
which is slightly more than the dispersion measured on this stack of 20 astronomical images.
  \label{fig:n17}}
\end{figure}

When the comparison is performed on the whole focal plane, 
a 10\% disagreement is found between the redistribution 
method and observations (figure~\ref{fig:d}).
Prediction for average X and Y brighter-fatter effect for the entire camera  are 
 0.022 pixel/100ke$^{-}$ in both directions, while it is found to be
0.025 pixel/100ke$^{-}$ in both directions from the observations.
This discrepancy does not come from the color correction 
that is applied to decorrelate the second moments of stars
from their color at fixed brightness. In the $r$-band, this correction is actually negligible 
(a 1$\cdot$10$^{-5}$ pixel/100ke$^{-}$ decreasing of the average slope).
The discrepancy is rather likely related to the non-linearity correction
that has to be applied to the images: the average of DECam flatfield departs from 
strict proportionality to the exposure time by $\sim$ 2\%.
If no corrections are applied, the average slope measurement 
is found to be 0.016 pixel/100ke$^{-}$. When the correction is 
 applied at the pixel level to linearize
DECam response to illumination, it shifts the average slope measurements
 to 0.025 pixel/100ke$^{-}$.\footnote{The non-linearity 
correction has no significant effect on the prediction.
When applied to the flatfield, it increases the average predicted slope
by $\cdot$10$^{-5}$ pixel/100ke$^{-}$.} 
With this level of non linearity, 
a careful handling is critical to obtain an accurate estimation of
the brighter-fatter effect.

\begin{figure}
\centering
\includegraphics[width=1.0\linewidth]{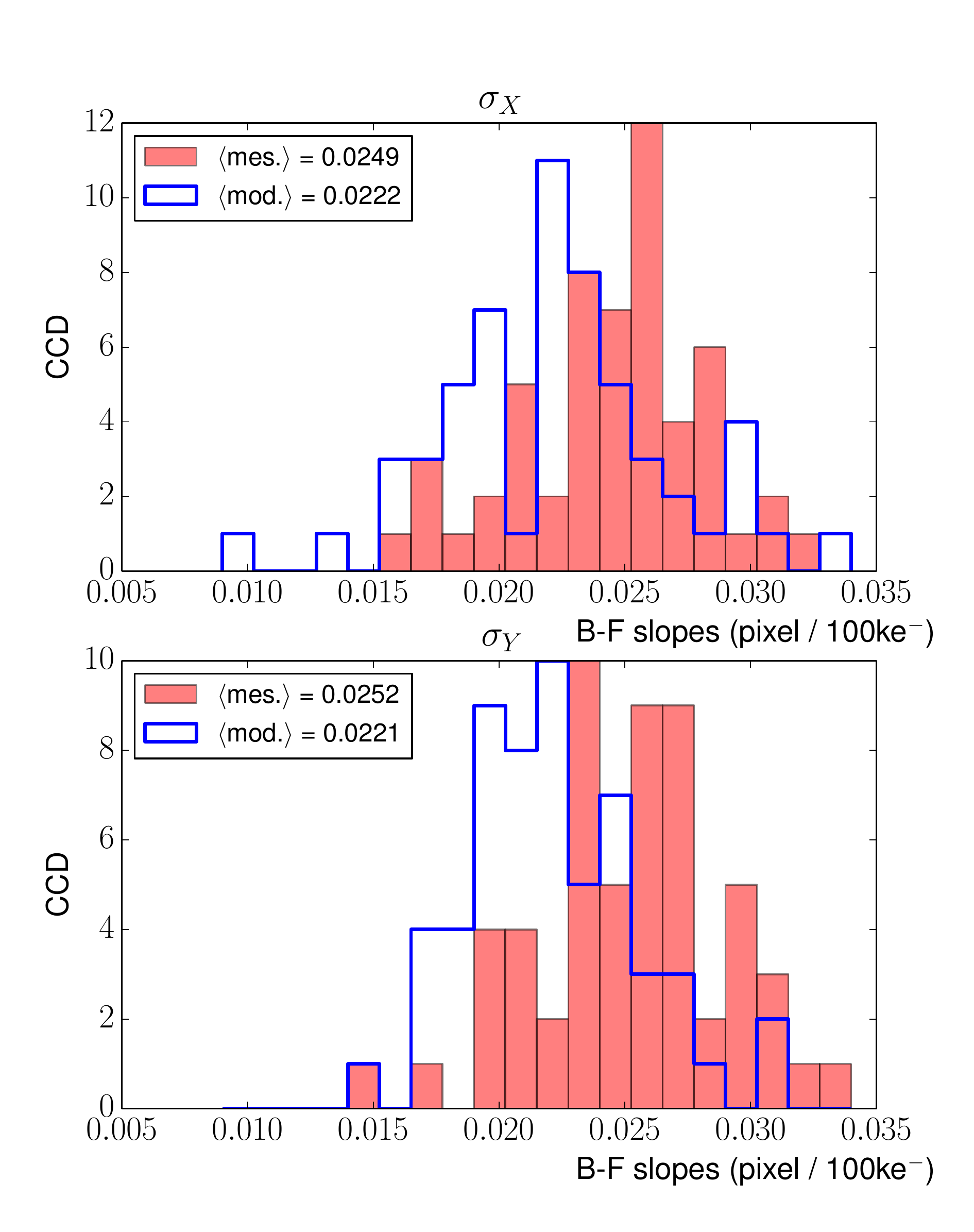}
\caption{Comparison of measured and simulated brighter-fatter slopes
  for DECam $r$-band. There are 57 CCDs over the 62 that are presented: we reject five CCDs 
   for which at least one of its amplifiers has a reduced $\chi^2$ for the linear fit 
   of the correlation $R_{1,1}$ that is above 3.
Astronomical images are being corrected at the pixel
  level to take non linearities detected from looking at
  flatfield mean fluxes versus exposure time into account. This increases the measured
  slopes by about 60 \%. While this correction is quite important, very
  small changes in the way non-linearities are fitted
  could account for the disagreement between measurements and model.
  \label{fig:d}}
\end{figure}

The mean prediction of stellar image broadening in the X and Y directions
 for the 36 MegaCam CCDs is presented in figure~\ref{fig:megbf} (shaded areas).
The figure gathers the correlations from all the CCDs so as to palliate the lack
of statistics coming from the small size of the frames in the flatfield
data set. The relative uncertainty is still quite large at almost 50 \%.
It is compared with a combination of all the epochs from the CFHTLS 
deep field (D1), which contains about 1.8M PSF estimations and which 
result in a high precision measurement. At this level, very small effects
show up, such as a small gap around 10 ke$^{-}$, the origin of which is not fully understood yet.
It should be remembered that the MegaCam brighter-fatter effect is already at a
few per mil level (while it is at a few percent level for thick CCDs) and that
the low-flux feature observed here is already at a 10$^{-4}$ level.

\begin{figure}
\centering
\includegraphics[width=0.9\linewidth]{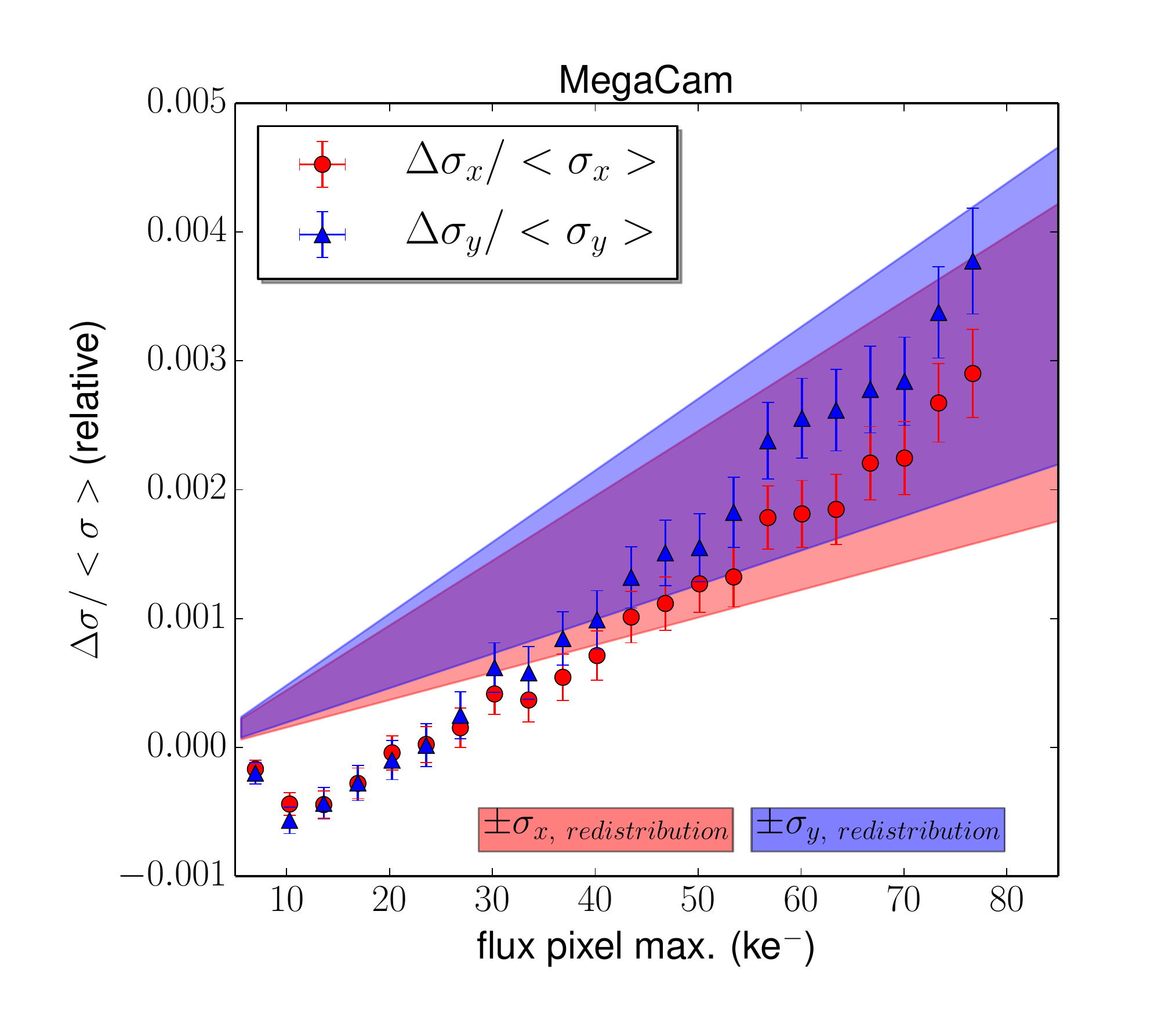}
\caption{Average brighter-fatter slope in the X and Y direction for the MegaCam mosaic. 
  The observations gather all the deep field (D1) $r$-band images of the CFHTLS 
($\approx$ 1.8 M stars). The redistribution prediction is based on the mean 
correlations of all the CCDs; this is necessary to palliate the lack 
of statistics given the small frame of the flatfields data set.
The prediction is compatible with the observations but with a significant uncertainty.
  \label{fig:megbf}}
\end{figure}

In this last section, we have presented the precision that can be achieved on predicting 
the brighter-fatter effect using our model. The approximations
that are made to tune the pixel effective size model are also presented, and their 
contributions to the systematics are evaluated. With the best quality data sample, 
as in the one from the CCD E2V-250, an agreement better than 5\% is 
found between the prediction and the measurement. We have also shown
that it can be used to reverse the redistribution of charges in pixels,
which allows to decorrelate the measurement of PSF widths and fluxes.
The precision that is reached is limited by the evaluation of the long range correlations.
This indicates that the quality of the collected PTCs is currently
the critical aspect to accurately correct for the broadening of point source image with increasing flux.

\section{Conclusion}

In this paper, we have studied two distinct effects that can
be observed in CCD data in detail.  The first one is the presence of long
distance pixel correlations within flatfield images, whose amplitude
increases linearly with flux (adding up to 8-18\% for 100 ke$^-$ levels). 
These correlations explain the sub-Poissonnian behavior of
the "Photon Transfer Curves" (PTC), which has been long noted on
multiple occasions by several authors.  The second effect is the
broadening of the PSF of stars with an increasing flux affecting
spot/star images (0.5-3\% for 100 ke$^-$ levels).  This so-called
"brighter-fatter effect" has been first observed on LSST sensor
candidates, and has since been confirmed by other teams.

Using three different data sets, collected with three
different types of sensors, we have shown that long distance pixel correlations and
PSF broadening are ubiquitous and seem to be a generic feature of CCD
devices. They not only affect the modern totally depleted thick sensors, 
which equip/will equip DECam and LSST, but they are also detectable on 
the thin CCDs which equip MegaCam.

We argue that pixel correlations and PSF broadening are deeply
connected.  Both effects can be described in terms of contrast
reduction as a function of the flux stored in the pixels. Hence, we
expect them to share a common physical explanation. We have developed
a physical model to explain our observations.  At the heart of this
model is the fact that the charges accumulated in each pixel produce
an alteration of the electric field within the CCD. Therefore, the pixel
boundaries which are set by the drift field evolve as the charges are
being collected.

To test this explanation, we have performed an electrostatic
simulation that calculates the alterations to the electric field
within the CCDs, as charges are being collected.  The outcome of the
simulation can be pictured as a shift of the pixel boundaries that
depends on its content and on the contents of the pixels in
its surroundings.  The model is able to reproduce (1) the long
distance correlations observed in flat fields, (2) the broadening of
the PSFs, and (3) the dependence of both on incoming flux and CCD operating
voltages.  In all cases, we
find that the amplitude of the effects that are predicted by the
simulation matches our observations.

Lastly, we have sketched a strategy to account for the brighter-fatter
effect in a data analysis pipeline. The main challenge here is the
precise characterization and modeling of the effect.  A possibility
would be to rely on extensive electrostatic simulations of the CCDs.  This
approach necessitates a detailed knowledge of the geometry of the
channel stops and of the doping concentration to accurately predict
the charge redistribution. This information, which is not usually
available, would be required with high precision to be able to 
predict both the increase of the diffusion and the evolution of the 
drift lines of the electrons.
The first effect is governed by the attenuation of the longitudinal field,
which is a short distance contribution that is only affecting adjacent pixels.
The latter contribution depends on both the longitudinal and the lateral electric field and extends 
up to several pixels separation.

Rather than a sophisticated simulation of charge trajectories, we propose an empirical model of dynamical charge transfers, which is tuned on the high quality
measurements of correlations in flatfields.  It is based on an algebra
that connects the displacement of each given pixel boundaries to the
correlations with the surrounding pixels measured on series of
flatfields.  The model is insensitive to the description of the
electrostatic within the CCDs.  The simultaneous determination of
boundary displacements within a given map of pixels correlation is
performed by setting (1) a limit condition on the correlations and (2) a
relation between orthogonal boundaries.
We find that this reverse redistribution method reduces the 
brighter-fatter effect
by more than one order of magnitude. 
We conclude that the recovery of 
a flux homothetic PSF is currently limited in precision by the quality 
of PTCs associated with a given CCD, from the estimation of the errors 
that are introduced by our assumptions and from the propagation of the
statistical uncertainties on correlation measurement. 

We have not discussed how these corrections can be implemented
practically in an image analysis pipeline.  There seem to be at least
two options.  A first possibility is to apply a reverse redistribution
correction at the pixel level, as done in this paper. Another is to incorporate the
effect directly in the PSF model.  The relative merits of these two
approaches depends on the science goal (e.g.,  whether it is measuring
fluxes or shapes). Deciding which is the best approach requires
additional work, and we defer this discussion to later papers.

\begin{acknowledgements}
The authors wish to thank Jean-Charles Cuillandre, Juan Estrada, Gary Bernstein
and Robert Lupton for useful discussions. The authors are grateful to the DES collaboration for acquiring and making available the DECam science verification data. We are also thankful to the LSST sensor team for providing the 
CCD E2V-250 data set.  
\end{acknowledgements}

\def\aap{{\em A\&A}}
\def\apj{ApJ}
\def\apjl{ApJ Lett.}
\def\apjs{ApJ Supp.}
\def\aj{AJ}
\def\prd{Phys. Rev. D}

\bibliographystyle{aa}
\bibliography{biblio}

\begin{thebibliography}{19}
\expandafter\ifx\csname natexlab\endcsname\relax\def\natexlab#1{#1}\fi

\bibitem[{Amara \& Réfrégier(2008)}]{Amara21112008}
Amara, A. \& Réfrégier, A. 2008, Monthly Notices of the Royal Astronomical
  Society, 391, 228

\bibitem[{{Antilogus} {et~al.}(2014){Antilogus}, {Astier}, {Doherty},
  {Guyonnet}, \& {Regnault}}]{BNL13}
{Antilogus}, P., {Astier}, P., {Doherty}, P., {Guyonnet}, A., \& {Regnault}, N.
  2014, ArXiv e-prints

\bibitem[{{Astier} {et~al.}(2013){Astier}, {El Hage}, {Guy}, {Hardin},
  {Betoule}, {Fabbro}, {Fourmanoit}, {Pain}, \& {Regnault}}]{Astier13}
{Astier}, P., {El Hage}, P., {Guy}, J., {et~al.} 2013, \aap, 557, A55

\bibitem[{{Bernstein} \& {Jarvis}(2002)}]{BernsteinJarvis02}
{Bernstein}, G.~M. \& {Jarvis}, M. 2002, \aj, 123, 583

\bibitem[{{Bernstein, G. et. al.}(2013)}]{berntein2013}
{Bernstein, G. et. al.} 2013, in Precision Astronomy with Fully Depleted CCDs,
  in Prep. Proc. of Cosmo 2013 at BNL

\bibitem[{{Betoule} {et~al.}(2014){Betoule}, {Kessler}, {Guy}, {Mosher},
  {Hardin}, {Biswas}, {Astier}, {El-Hage}, {Konig}, {Kuhlmann}, {Marriner},
  {Pain}, {Regnault}, {Balland}, {Bassett}, {Brown}, {Campbell}, {Carlberg},
  {Cellier-Holzem}, {Cinabro}, {Conley}, {D'Andrea}, {DePoy}, {Doi}, {Ellis},
  {Fabbro}, {Filippenko}, {Foley}, {Frieman}, {Fouchez}, {Galbany}, {Goobar},
  {Gupta}, {Hill}, {Hlozek}, {Hogan}, {Hook}, {Howell}, {Jha}, {Le Guillou},
  {Leloudas}, {Lidman}, {Marshall}, {M{\"o}ller}, {Mour{\~a}o}, {Neveu},
  {Nichol}, {Olmstead}, {Palanque-Delabrouille}, {Perlmutter}, {Prieto},
  {Pritchet}, {Richmond}, {Riess}, {Ruhlmann-Kleider}, {Sako}, {Schahmaneche},
  {Schneider}, {Smith}, {Sollerman}, {Sullivan}, {Walton}, \&
  {Wheeler}}]{Betoule14}
{Betoule}, M., {Kessler}, R., {Guy}, J., {et~al.} 2014, ArXiv e-prints

\bibitem[{Borgeaud {et~al.}(2000)Borgeaud, Gallais, Boulade, Carton, Gros,
  de~Kat, Lee, \& Nemee}]{doi:10.1117/12.395493}
Borgeaud, P., Gallais, P., Boulade, O., {et~al.} 2000, 40 CCDs of the MegaCam
  wide-field camera: procurement, testing, and first laboratory results

\bibitem[{{Boulade} {et~al.}(2003){Boulade}, {Charlot}, {Abbon}, {Aune},
  {Borgeaud}, {Carton}, {Carty}, {Da Costa}, {Deschamps}, {Desforge},
  {Eppell{\'e}}, {Gallais}, {Gosset}, {Granelli}, {Gros}, {de Kat}, {Loiseau},
  {Ritou}, {Rouss{\'e}}, {Starzynski}, {Vignal}, \& {Vigroux}}]{Boulade2003}
{Boulade}, O., {Charlot}, X., {Abbon}, P., {et~al.} 2003, in Society of
  Photo-Optical Instrumentation Engineers (SPIE) Conference Series, Vol. 4841,
  Society of Photo-Optical Instrumentation Engineers (SPIE) Conference Series,
  ed. {M.~Iye \& A.~F.~M.~Moorwood}, 72--81

\bibitem[{{Chang} {et~al.}(2012){Chang}, {Kahn}, {Jernigan}, {Peterson},
  {AlSayyad}, {Ahmad}, {Bankert}, {Bard}, {Connolly}, {Gibson}, {Gilmore},
  {Grace}, {Hannel}, {Hodge}, {Jee}, {Jones}, {Krughoff}, {Lorenz}, {Marshall},
  {Marshall}, {Meert}, {Nagarajan}, {Peng}, {Rasmussen}, {Shmakova},
  {Sylvestre}, {Todd}, \& {Young}}]{Chang12}
{Chang}, C., {Kahn}, S.~M., {Jernigan}, J.~G., {et~al.} 2012, ArXiv e-prints

\bibitem[{{Downing} {et~al.}(2006){Downing}, {Baade}, {Sinclaire}, {Deiries},
  \& {Christen}}]{Downing06}
{Downing}, M., {Baade}, D., {Sinclaire}, P., {Deiries}, S., \& {Christen}, F.
  2006, in Society of Photo-Optical Instrumentation Engineers (SPIE) Conference
  Series, Vol. 6276, Society of Photo-Optical Instrumentation Engineers (SPIE)
  Conference Series

\bibitem[{{Downing} \& {Sinclaire}(2013)}]{Downing13}
{Downing}, M. \& {Sinclaire}, P. 2013, in Scientific Detector Workshop,
  Florence, Italy

\bibitem[{{Estrada, J. et. al. }(2010)}]{Estrada2010}
{Estrada, J. et. al. }. 2010, in Proc. of SPIE - Ground-based and Airborne
  Instrumentation for Astronomy III, Vol. 7735, Society of Photo-Optical
  Instrumentation Engineers (SPIE) Conference Series, ed. {Ian S. McLean,
  Suzanne K. Ramsay, Hideki Takam}, 72--81

\bibitem[{{Holland, S. et. al}(2003)}]{holland2003}
{Holland, S. et. al}. 2003, IEEE, 50

\bibitem[{{Holland, S. et. al.}(2013)}]{holland13}
{Holland, S. et. al.} 2013, in Precision Astronomy with Fully Depleted CCDs, in
  Prep. Proc. BNL workshop Cosmo 2013

\bibitem[{{Kent}(1973)}]{Kent73}
{Kent}, W.~H. 1973, The Bell System Technical Journal, 52

\bibitem[{{Laureijs} {et~al.}(2011){Laureijs}, {Amiaux}, {Arduini}, {Augueres},
  {Brinchmann}, {Cole}, {Cropper}, {Dabin}, {Duvet}, {Ealet}, \&
  et~al.}]{EuclidRB}
{Laureijs}, R., {Amiaux}, J., {Arduini}, S., {et~al.} 2011, ArXiv e-prints

\bibitem[{{Ma} {et~al.}(2014){Ma}, {Shang}, {Wang}, {Hu}, {Liu}, \& {Wei}}]{ma}
{Ma}, B., {Shang}, Z., {Wang}, L., {et~al.} 2014, ArXiv e-prints

\bibitem[{{Marshall} {et~al.}(2013){Marshall}, {Rheault}, {DePoy}, {Prochaska},
  {Allen}, {Behm}, {Martin}, {Veal}, {Villanueva}, {Williams}, \&
  {Wise}}]{marshall2013}
{Marshall}, J.~L., {Rheault}, J.-P., {DePoy}, D.~L., {et~al.} 2013, ArXiv
  e-prints

\bibitem[{{Melchior} {et~al.}(2014){Melchior}, {Suchyta}, {Huff}, {Hirsch},
  {Kacprzak}, {Rykoff}, {Gruen}, {Armstrong}, {Bacon}, {Bechtol}, {Bernstein},
  {Bridle}, {Clampitt}, {Honscheid}, {Jain}, {Jouvel}, {Krause}, {Lin},
  {MacCrann}, {Patton}, {Plazas}, {Rowe}, {Vikram}, {Wilcox}, {Young}, {Zuntz},
  {Abbott}, {Abdalla}, {Allam}, {Banerji}, {Bernstein}, {Bernstein}, {Bertin},
  {Buckley-Geer}, {Burke}, {Castander}, {da Costa}, {Cunha}, {Depoy}, {Desai},
  {Diehl}, {Doel}, {Estrada}, {Evrard}, {Fausti Neto}, {Fernandez}, {Finley},
  {Flaugher}, {Frieman}, {Gaztanaga}, {Gerdes}, {Gruendl}, {Gutierrez},
  {Jarvis}, {Karliner}, {Kent}, {Kuehn}, {Kuropatkin}, {Lahav}, {Maia},
  {Makler}, {Marriner}, {Marshall}, {Merritt}, {Miller}, {Miquel}, {Mohr},
  {Neilsen}, {Nichol}, {Nord}, {Reil}, {Roe}, {Roodman}, {Sako}, {Sanchez},
  {Santiago}, {Schindler}, {Schubnell}, {Sevilla-Noarbe}, {Sheldon}, {Smith},
  {Soares-Santos}, {Swanson}, {Sypniewski}, {Tarle}, {Thaler}, {Thomas},
  {Tucker}, {Walker}, {Wechsler}, {Weller}, \& {Wester}}]{2014arXiv1405.4285M}
{Melchior}, P., {Suchyta}, E., {Huff}, E., {et~al.} 2014, ArXiv e-prints

\end{thebibliography}

\appendix

\section{Recovering the Poisson  variance of flatfields}
\label{recovar}
We derive how one can evaluate the Poisson variance of flatfields
from the measurement of the actual variance and covariances here. We assume
that the observed flatfields are a perturbed realization of ideal
flatfields ,which would have independent pixels in the absence of
perturbation. 

In \cite{Downing06}, it is proposed to sum the covariances of the
perturbed images in order to recover the variance of the unperturbed
image but without justifying this procedure.  Note that the sum of
covariances of the unperturbed image is equal to its variance, because
covariances are zero there by hypothesis.  So, in order to justify the
procedure, we have to show that the perturbations conserve the sum of
covariances, which is the integral of the correlation function.

Let us define
\begin{align}
Q'_{ij} &= Q_{ij} + \delta Q_{ij}  \nonumber \\
C_{kl} &\equiv \frac{1}{N}\sum_{ij} Q'_{i,j} Q'_{i+k,j+l} - \mu^2  \; , \nonumber
\end{align}
where $Q$ and $Q'$ refer to the unperturbed and perturbed images respectively; 
$\mu$ is their (common) average; and $N$ is the number of pixels in the sum.

We have
\begin{align}
\sum_{kl} C_{kl} &= \frac{1}{N} \sum_{kl}\sum_{ij} Q'_{i,j} Q'_{i+k,j+l} - \mu^2 \nonumber \\
               &= Var(Q) + \frac{2}{N} \sum_{kl}\sum_{ij} \delta Q_{i,j} Q_{i+k,j+l} + O(\delta Q^2) \nonumber \\
               &= Var(Q) + \frac{2}{N} \sum_{ij} \delta Q_{i,j} \sum_{kl} Q_{k,l} + O(\delta Q^2)  \; , \nonumber
\end{align}
where the last transformation is only exact for unbounded sums. Since
the perturbations exactly conserve charge, we have $\sum_{ij} \delta
Q_{i,j} \equiv 0$, and hence the perturbation preserves the sum of the
correlations to first order of perturbations. We note that charge
conservation is almost exact over small image patches and hence,
ignoring boundary effects in the algebraic argument makes sense.  This
property allows us to recover the Poisson variance $V(Q)$ 
by summing all correlations $\sum_{kl} C_{kl}$. It should be emphasized
that, only a subset 
of the $C_{kl}$ are summed (for instance, 81 coefficients are summed 
in figure \ref{fig:varprof})in practice, which always results in an underestimation of 
the Poisson variance (about 0.5\% underestimation on these examples).

In \cite{Downing06}, it is also proposed to rebin the image to
recover the linearity of the PTC. Rebinning still leaves some
correlation between larger pixels and hence cannot exactly provide the
variance of independent pixels. If one rebins an image into $n$ times
bigger pixels (in each direction), the expression for the variance of
the ``big'' pixels divided by $n^2$ is a linear combination
of covariances, with coefficients smaller than 1, and which approaches 1 as one rebins
into pixels of increasing size. For example, the covariance of nearest
neighbors is multiplied by $(n-1)/n$ for the variance of a rebinned
image (by $n$ in both direction). Rebinning provides
an easy and fair approximation of the Poisson variance but is not
exact. One has to rebin by typically 10 to approach a 1 \% accuracy
of the variance around the full well.

\end{document}